\documentclass[12pt,a4paper]{article}
\usepackage[utf8]{inputenc}
\usepackage{epsf}
\usepackage{latexsym,amssymb,euscript}
\usepackage[dvips]{graphicx}
\usepackage[numbers,sort&compress]{natbib}
\usepackage{amsmath}
\usepackage{nicefrac}
\usepackage{slashed}
\usepackage{booktabs}
\usepackage{hyperref}
\usepackage{braket}
\usepackage{chngcntr}
\usepackage{bbold}
\usepackage{graphics}
\usepackage{graphicx}
\usepackage{mciteplus}
\usepackage{caption}
\usepackage{subcaption}
\usepackage{pdfpages}
\usepackage[titletoc]{appendix}
\graphicspath{{./figures/}}
\hypersetup{
 linktocpage = true,
 urlcolor = teal,
 colorlinks = true,
 linkcolor = teal,
 anchorcolor = teal,
 citecolor = teal,
 pdfstartview = {XYZ null null 1.25} 
           }
\usepackage[left=1.9cm, right=1.9cm]{geometry}
\usepackage{pstricks}
\usepackage{color}
\usepackage{float}
\usepackage{academicons}
\definecolor{orcidlogocol}{HTML}{A6CE39}
\usepackage{fancyhdr}
\newcommand{\drv}{{\rm d}}

\newcommand{\as}{\alpha_s}

\newcommand{\MSb}{\overline{\rm MS}}
\newcommand{\CnLLA}{{\cal C}_n^{\rm LLA}}
\newcommand{\CnNLA}{{\cal C}_n^{\rm NLA}}

\newcommand{\DY}{\Delta Y}
\newcommand{\JPsi}{J/\psi}
\newcommand{\Yps}{\Upsilon}
\newcommand{\Q}{{\cal Q}}

\newcommand{\tcite}[1]{~\cite{#1}}
\newcommand{\tref}[1]{~\ref{#1}}
\newcommand{\eref}[1]{~\eqref{#1}}

\newcommand{\tarr}{
\begin{array}}
\newcommand{\earr}{\end{array}}

\newcommand{\orcidFGC}{\href{https://orcid.org/0000-0003-3299-2203}{\includegraphics[scale=0.1]{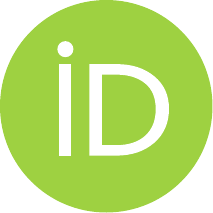}}}

\newcommand{\orcidMF}{\href{https://orcid.org/0000-0002-2408-2210}{\includegraphics[scale=0.1]{figures/logo-orcid.pdf}}}

\begin{document}

\begin{titlepage}

\begin{center}
  {\LARGE \bf Diffractive semi-hard production of a $\JPsi$ or a $\Yps$ \\ from single-parton fragmentation plus a jet \vskip.025cm in hybrid factorization}
\end{center}

\vskip 0.35cm

\centerline{
Francesco Giovanni~Celiberto$^{\;1,2,3\;\dagger}$ \orcidFGC \, and
Michael~Fucilla$^{\;4,5,6\;\ddagger}$ \orcidMF
}

\vskip .4cm

\centerline{${}^1$ {\sl European Centre for Theoretical Studies in Nuclear Physics and Related Areas (ECT*),}}
\centerline{\sl I-38123 Villazzano, Trento, Italy}
\vskip .18cm
\centerline{${}^2$ {\sl Fondazione Bruno Kessler (FBK), 
I-38123 Povo, Trento, Italy} }
\vskip .18cm
\centerline{${}^3$ {\sl INFN-TIFPA Trento Institute of Fundamental Physics and Applications,}}
\centerline{\sl I-38123 Povo, Trento, Italy}
\vskip .18cm
\centerline{${}^4$ {\sl Dipartimento di Fisica, Universit\`a della Calabria,}}
\centerline{\sl I-87036 Arcavacata di Rende, Cosenza, Italy}
\vskip .18cm
\centerline{${}^5$ {\sl Istituto Nazionale di Fisica Nucleare, Gruppo collegato di Cosenza,}}
\centerline{\sl I-87036 Arcavacata di Rende, Cosenza, Italy}
\vskip .18cm
\centerline{${}^6$ { \sl Université Paris-Saclay, CNRS, IJCLab, 91405 Orsay, France}}
\vskip 0.55cm

\begin{abstract}
\vspace{0.25cm}
\hrule \vspace{0.25cm}
We investigate the inclusive hadroproduction of a heavy quarkonium ($\JPsi$ or $\Yps$), in association with a light-flavored jet, as a testing ground for the semi-hard regime of QCD.
Our theoretical setup is the hybrid high-energy and collinear factorization, where the standard collinear approach is supplemented by the $t$-channel resummation of leading and next-to-leading energy logarithms \emph{à la} BFKL.
We present predictions for rapidity and azimuthal-angle differential distributions, hunting for stabilizing effects of the high-energy series under higher-order corrections.
Our reaction represents an additional channel to test the production mechanisms of quarkonia at high energies and large transverse momenta and to possibly shed light on the transition region from heavy-quark pair production to single-parton fragmentation of $\JPsi$ and $\Yps$ states.
\vspace{0.25cm} \hrule
\vspace{0.25cm}
{
 \setlength{\parindent}{0pt}
}
\end{abstract}

\vfill
$^{\dagger}${\it e-mail}:
\href{mailto:fceliberto@ectstar.eu}{fceliberto@ectstar.eu}

$^{\ddagger}${\it e-mail}:
\href{mailto:michael.fucilla@unical.it}{michael.fucilla@unical.it}




\end{titlepage}

\section{Hors d'{\oe}uvre}
\label{sec:intro}

The study of production mechanisms of heavy-quark bound states in high-energy collisions has always played a key role in widening our knowledge of the dynamics of strong interactions.
A relevant class of heavy-flavored hadrons is represented by mesons whose lowest Fock state is composed of a heavy quark and the corresponding antiquark. These mesons are called heavy \emph{quarkonia}, or simply quarkonia.
The era of quarkonium physics began with the \emph{November Revolution}, when on 11 November 1974 the discovery of a new particle with a mass of about 3.1 GeV and the same quantum numbers of the photon was announced. That particle is the $\JPsi$ vector meson and owes its name to the two groups that simultaneously observed it, namely Collaboration leaded by B.~Richter\tcite{SLAC-SP-017:1974ind} at the Stanford Linear Accelerator Center (SLAC) and one headed by S.~Ting\tcite{E598:1974sol} at the Brookhaven National Laboratory (BNL).
Seven days after the SLAC and BNL announcements, the $\JPsi$ discovery was confirmed by the ADONE experiment in Frascati, under the direction of G.~Bellettini\tcite{Bacci:1974za}.
The hadronic nature of the $\JPsi$ was proven by the cross-section ratio between hadrons and $\mu^+\mu^-$ in inclusive $e^+e^-$ inclusive annihilations that turned out to be much higher at the resonance, this meaning that $\JPsi$-like particles directly decay into hadronic states.
This provided us with a proof of the existence of the charm-quark flavor and, more in general, with a clear evidence that quarks are real constituents and not just mathematical artifacts. The existence of the charm quark was first proposed in 1964 by Bjorken and Glashow\tcite{Bjorken:1964bo}, but there were few experimental evidences. In 1970, with the proposal of the Glashow--Iliopoulos--Maiani (GIM) mechanism~\cite{Glashow:1970gi}, the existence of a new quark species, the charm one, became a necessary ingredient to explain why the flavor-changing neutral currents (FCNCs) are suppressed in loop diagrams.

Quarkonium studies represented one of the first testing grounds for the manifestation of fundamental properties of Quantum ChromoDynamics (QCD), such as the \emph{asymptotic freedom}. Indeed, with the discovery of the $\psi(2S)$, the first radial excited state of the $\JPsi$, it became clear that strong interactions decrease at short distances, resulting in a Coulomb-like binding potential with a confinement part.
The $\JPsi$ was the first \emph{charmonium} state to be discovered. The name charmonium was given in analogy with a $e^+e^-$ bound pair, that is a \emph{positronium}, and represents a family of hadrons which comprehends all the $c \bar c$ meson states. 
In the years following the November Revolution other charmonia were observed (as the $P$-wave $\chi_c$ and the pseudoscalar $\eta_c$), as well as open-charm $D$ mesons\tcite{Wiss:1976gd}.
In 1977 the first \emph{bottomonium} state was observed. It was called $\Upsilon$, and it is a vector meson equivalent to the $\JPsi$, but with bottom quarks\tcite{Herb:1977ek}. It was followed by the discovery of excited $b \bar b$ states (as the $\Upsilon(2S)$) and of open-bottom $B$ mesons\tcite{CLEO:1980oyr}.

Despite the ease of performing experimental studies on quarkonia, especially on the vector mesons which can be easily observed in lepton-lepton reactions, the phenomenological description of their production rates still remains a challenge and different models for the quarkonium production mechanism have been proposed so far (see Refs.\tcite{Lansberg:2019adr,Scarpa:2020sdy} for further details).
An early-day model, known as Color Evaporation Model (CEM)\tcite{Fritzsch:1977ay,Halzen:1977rs}, postulates that the quarkonium-constituent $(Q \bar Q)$-pair undergoes a large number of soft interactions that make it a color singlet hadronizing into the observed quarkonium. Soft interactions lead to a complete color decorrelation between the produced partons and the pair, so that even a single gluon can evolve into the final hadron. The main limit of the CEM is the inability to afford a different description for production rates of distinct quarkonia, as the ones of $\JPsi$ and $\chi_c$\tcite{Lansberg:2006dh,Brambilla:2010cs}.

A second attempt at describing quarkonium production relies on the so-called Color Singlet Mechanism (CSM)\tcite{Einhorn:1975ua,Chang:1979nn,Berger:1980ni,Baier:1981uk}. It is based on the assumption, opposite to the CEM one, that the $(Q \bar Q)$-pair does not evolve due to the suppression of gluon emissions with powers of $\alpha_s(m_Q)$, where $m_Q$ is the heavy-quark mass.
Since the quark spin and color remain unaltered during the hadronization, and since the quarkonium must be color-neutral, the $(Q \bar Q)$-pair needs to be generated in a color-singlet state. 
Furthermore, since the quarkonium mass, $m_{\cal Q}$, turns out to be slightly larger than $2 m_Q$, the two constituent quarks can be assumed to be almost at rest in the hadron frame, namely with zero
relative velocity, $v_{\cal Q}$, and the only non-perturbative contribution to the production mechanism is a non-relativistic Schr\"odinger wave function at the origin, ${\cal R}_{\cal Q}(0)$.
The strength of the CSM consists in its relatively high predictive power. Indeed, the value of the radial wave function at the origin is the only free parameter which can be extracted from measurements of  quarkonium leptonic-decay widths.
However, the CSM suffers from infrared singularities arising at the next-to-leading order (NLO) in $P$-wave decay channels \tcite{Barbieri:1976fp,Bodwin:1992ye}.

To go beyond the CSM approximation of $(Q \bar Q)$ static quantum numbers, it was proposed that higher Fock states (\emph{e.g.}, $|Q \bar Q g \rangle$), contribute to the quarkonium production and  might remove the NLO singularity.In a $|Q \bar Q g \rangle$ state the quark-pair is in a color-octet (CO) configuration. More generally, the physical quarkonium is given as a linear combination of all possible Fock states. All contributions are organized as a double expansion in powers of $\alpha_s$ and $v_{\cal Q}$, of which the CSM represents the leading term in $v_{\cal Q}$ for $S$-wave channels (for higher waves this contribution may be zero). This double expansion is the building block of an effective field theory (EFT) known as Non-Relativistic QCD (NRQCD)\tcite{Caswell:1985ui,Thacker:1990bm,Bodwin:1994jh}.
Within this approach cross sections for quarkonium productions take the form of a sum of partonic cross sections creating a given Fock state, each of them being multiplied by a Long-Distance
Matrix Element (LDME) encoding the non-perturbative hadronization.
Although affording a very appealing solution to the quarkonium production puzzle, the NRQCD introduces many new non-perturbative parameters that makes non-trivial the comparison with data.

Another relevant issue comes from the fact that a quarkonium state can be produced via the decay or the de-excitation of a heavier hadron.
As an example, a $\JPsi$ can be produced from the decay of $B$~meson\tcite{Halzen:1984rq}. This channel is \emph{non-prompt}, since there is a visible time delay due to the decay.
Some experimental studies have managed to isolate the non-prompt contribution by detecting the distance between the creation vertex of the $B$ particle and its decay or de-excitation to the quarkonium state. This procedure permits us to get the \emph{prompt} component.
Within the prompt channel, a charmed quarkonium, say the $J/\psi$, can be still generated from the decay of a $\chi_c$, or from the de-excitation of a $\psi(2S)$. This is known as \emph{indirect} production component.
Conversely, the fraction of hard-scattering produced quarkonia is known as \emph{direct} component and is generally not accessible alone, but can be accessed by subtraction only.

The models described above rely on the assumption that the dominant production mechanisms is the \emph{short-distance} production of a $(Q \bar Q)$ pair in the hard scattering which then hadronizes into the detected quarkonium.
Since the quark and the antiquark are created with a relative transverse separation of order
$1/p_T$, this channel is expected to be suppressed at large $p_T$. According to the short-distance mechanism, the $(Q \bar Q)$ pair is created with a transverse separation of order $1/Q$, with $Q$ the characteristic energy scale of the process\tcite{Mangano:1995yd}.
In the large transverse-momentum range one has $Q \sim p_T \gg m_Q$.
Thus, exchanges of particles having virtualities of the same order of $p_T$ quenches the probability amplitude, since they are related to scattering of particles in a quite small volume, $1/p_T^3$\tcite{Braaten:1996pv,Artoisenet:2009zwa}.

\begin{figure*}[!t]
\centering
\includegraphics[width=0.45\textwidth]{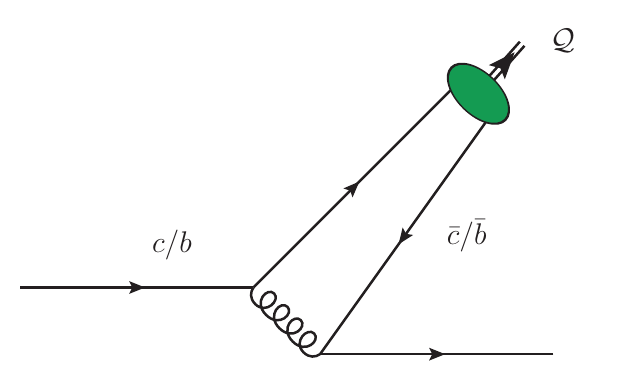}
\hspace{0.50cm}
\includegraphics[width=0.45\textwidth]{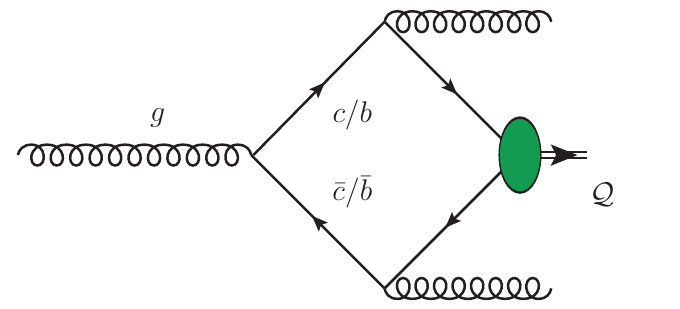}

\caption{Left: one of the leading diagrams contributing to the heavy-quark fragmentation to a $^3S_1^{(1)}$ vector quarkonium $\cal Q$ at order $\alpha_s^2$.
Right: one of the leading diagrams contributing to the gluon fragmentation to $^3S_1^{(1)}$ vector quarkonium $\cal Q$ at order $\alpha_s^3$.
The green blob denotes the corresponding non-perturbative NRQCD LDME.}
\label{fig:FF_diagrams}
\end{figure*}

In Ref.\tcite{Braaten:1993rw} it was highlighted that in this kinematic regime an additional mechanism comes into play, namely the \emph{fragmentation} of a large-$p_T$ parton (gluon or quark) which afterwards decays into the quarkonium state plus other partons.
The hadronization process is described by a set of collinear Fragmentation Functions (FFs) which evolve according to the  Dokshitzer--Gribov--Lipatov--Altarelli--Parisi (DGLAP) equation\tcite{Gribov:1972ri,Gribov:1972rt,Lipatov:1974qm,Altarelli:1977zs,Dokshitzer:1977sg}. NRQCD provides us a way to perturbatively calculate these FFs.
Here, the $(Q \bar Q)$ pair is produced with a separation of order $1/m_Q$. Although the fragmentation mechanism is often of higher perturbative order with respect to the short-distance one, it is enhanced by $(Q/m_Q)^2$. Therefore, it prevails at high energies, $Q \gg m_Q$. Fig.\tref{fig:FF_diagrams} shows a selection of leading Feynman diagrams contributing to vector-quarkonium fragmentation.
%
The NRQCD calculation of the gluon FF to a $S$-wave quarkonium in the CSM was completed in Ref.\tcite{Braaten:1993rw} for $\JPsi$ and $\eta_c$ at leading order (LO) and in Ref.\tcite{Artoisenet:2014lpa} for $\eta_c$ at NLO.
The corresponding results for the FF depicting the $c \to J/\psi+c+X$ were obtained in Ref.\tcite{Braaten:1993mp} at LO and in Ref.\tcite{Zheng:2019dfk} at NLO.
LO phenomenological studies on unveiling the transition region between the short-distance and the fragmentation mechanisms were performed in Refs.\tcite{Doncheski:1993xm,Braaten:1994xb,Cacciari:1994dr,Cacciari:1995yt,Cacciari:1995fs}.
FFs from a heavy-quark pair in the $S$- and $P$-wave channels were investigated in Ref.\tcite{Ma:2013yla} and\tcite{Ma:2014eja}, respectively.
In those works next-to-leading power studies in $p_T$ which have shown that the leading-power single-parton fragmentation is relevant at larger $p_T$-values, rather than what initially found in the 90’s.
A next-to-NLO factorization analysis on quarkonium fragmentation functions can be found in Refs.\tcite{Nayak:2005rw,Nayak:2005rt}.
FFs for polarized quarkonia were studied in Refs.\tcite{Falk:1993rj,Chen:1993ii,Cho:1994gb,Kang:2011mg,Kang:2014pya,Ma:2015yka} (see Ref.\tcite{Lansberg:2008gk} for a discussion).
Ref.\tcite{Kang:2014tta} contains analyses on quarkonium factorization at evolution at large transverse momenta.

Despite the ambiguities in the description of their production mechanisms and in the correct interpretation of experimental data, quarkonium emissions are considered excellent \emph{tools} to investigate properties of strong interactions.
First studies of quarkonium production in collinear factorization and with NLO accuracy can be found, \emph{e.g.}, in Refs.\tcite{Mangano:1991jk,Kuhn:1992qw,Kramer:1995nb,Petrelli:1997ge,Maltoni:1997pt,Klasen:2004tz,Gong:2008ft,Maltoni:2006yp,Lansberg:2013qka,Lansberg:2017ozx}. Refs.\tcite{Lansberg:2016rcx,Lansberg:2020rft} contain NLO analyses on the transverse-momentum spectrum within CEM. Investigations on quarkonium polarization in NRQCD were done in Refs.\tcite{Chao:2012iv,Shao:2014fca,Shao:2014yta}.
Multi-Parton Interactions (MPI) effects in $\JPsi$ production were studied in Refs.\tcite{Kom:2011bd,Lansberg:2014swa}.
The impact of the nuclear modification of the gluon density on quarkonium was recently assessed\tcite{Lansberg:2016deg,Kusina:2017gkz,Kusina:2020dki}.
Advances in the automation of NRQCD calculations for quarkonium emissions are reported in Refs.\tcite{Artoisenet:2007qm,Shao:2012iz,Shao:2015vga,Shao:2018adj}.
Issues in the large-$p_T$ description of $\JPsi$ production at NLO were attributed to an over-subtraction in the factorization  the collinear divergences inside collinear Parton Distribution Functions (PDFs)\tcite{Flore:2020jau,Lansberg:2020ejc,ColpaniSerri:2021bla}, and a possible solution was found in the inclusion of high-energy effects on top of NLO calculations\tcite{Lansberg:2021vie}.
The nature of hidden-charm and bottom \emph{tetra-} and \emph{penta-quarks} was investigated\tcite{Ferretti:2018ojb,Ferretti:2020ewe} via the pioneering \emph{hadro-quarkonium} picture\tcite{Dubynskiy:2008mq,Guo:2009id}.
Efforts in obtaining tetra-quark FFs from NRQCD-inspired calculations have been recently made\tcite{Feng:2020riv,Nejad:2021mmp}.

Semi-inclusive quarkonium-detection processes are widely recognized as golden channels to access the three-dimensional structure of hadrons via \emph{transverse-momentum-dependent} gluon distributions\tcite{Mulders:2000sh,Meissner:2007rx,Lorce:2013pza,Boer:2016xqr}. Recent phenomenological analyses of gluon TMDs through observables sensitive to quarkonium production in electron-proton and proton-proton collisions were can be found in Refs.\tcite{DAlesio:2019qpk,Boer:2020bbd,Bacchetta:2018ivt,Boer:2021ehu,DAlesio:2021yws} and\tcite{denDunnen:2014kjo,Lansberg:2017dzg,Scarpa:2019fol,DAlesio:2020eqo}, respectively.\footnote{See Ref.\tcite{Boer:2016bfj} for an overview of phenomenological opportunities.}
Studies on TMD factorization for heavy-quarkonium emissions were recently conducted in a Soft and Collinear Effective Theory (SCET) \cite{Echevarria:2019ynx,Fleming:2019pzj}, and then extended to di-jet and heavy-meson pair emissions in lepton-proton scatterings\tcite{delCastillo:2020omr}. A SCET approach was employed to access the jet substructure and, more in particular, to define fragmenting jet functions, needed to describe the production of quarkonia inside jets\tcite{Procura:2009vm,Baumgart:2014upa,Bain:2016clc,Bain:2016rrv,Bain:2017wvk}. The $p_T$-spectrum of quarkonium production from single light-parton fragmentation was recently investigated at the hands of SCET and NRQCD\tcite{Echevarria:2020qjk}.
Phenomenological analyses of quarkonium emissions at new-generation colliders via a novel determination of polarized gluon TMDs\tcite{Bacchetta:2020vty,Celiberto:2021zww,Bacchetta:2021oht,Bacchetta:2021lvw,Bacchetta:2021twk,Bacchetta:2022esb,Bacchetta:2022nyv} are underway.

Quarkonium studies at low-$x$ are relevant to observe the growth with energy of cross sections\tcite{Bautista:2016xnp} predicted by the linear Balitsky--Fadin--Kuraev--Lipatov (BFKL) evolution\tcite{Fadin:1975cb,Kuraev:1976ge,Kuraev:1977fs,Balitsky:1978ic}, as well as the onset of possible the non-linear \emph{gluon-recombination} effects\tcite{Garcia:2019tne} leading to the \emph{saturation} pattern.
In Ref.\tcite{Kang:2013hta} the emission of quarkonia in high-energy proton-nucleus scatterings was studied by making use of NRQCD and by accounting for small-$x$ evolution and multiple-scattering effects within the Color Glass Condensate (CGC) formalism, a semi-classical EFT for saturation (see Refs.\tcite{Gelis:2010nm,Kovchegov:2012mbw} for a comprehensive overview).
A selection of recent results on forward and central quarkonium production in (ultra-peripheral) hadron and lepton-hadron collisions can be found in Refs.\tcite{GayDucati:2013sss,Kniehl:2016sap,GayDucati:2016ryh,Goncalves:2017wgg,Cisek:2017ikn,Cepila:2017nef,Maciula:2018bex,Goncalves:2018blz,Babiarz:2019sfa,Babiarz:2019mag,Goncalves:2019txs,Babiarz:2020jkh,Babiarz:2020jhy,Kopp:2020pjk,Guzey:2020ntc,Xie:2021seh,Jenkovszky:2021sis}.
Combining the information coming from single-forward emissions of quarkonia with the one encoded in single-forward detections of light vector mesons~\cite{Anikin:2011sa,Besse:2013muy,Bolognino:2018rhb,Bolognino:2018mlw,Bolognino:2019bko,Bolognino:2019pba,Celiberto:2019slj,Bolognino:2021niq,Bolognino:2021gjm,Bolognino:2022uty} and in the forward Drell--Yan di-lepton reaction\tcite{Motyka:2014lya,Brzeminski:2016lwh,Celiberto:2018muu} will permits us to perform an extensive scan of the intersection kinematic range between TMD and high-energy dynamics. 
In particular, the main outcome of Refs.\tcite{Bolognino:2018rhb,Bolognino:2021niq} is that the forward polarized $\rho$-meson leptoproduction at HERA and the EIC act as an excellent probe channel for the proton content at small-$x$, the corresponding $p_T$-factorized cross sections being sensitive to a small transverse-size dipole-scattering mechanism. 
On the other side, forward tags of excited states, such as $\psi(2S)$ and $\Yps(2S)$ mesons, could lead to quite larger sizes of dipoles\tcite{Suzuki:2000az,Cepila:2019skb,Hentschinski:2020yfm}, this calling for a TMD-inspired enhancement of the pure small-$x$ description.

In this work we turn our attention to the inclusive hadroproduction of heavy vector mesons at high-energies, as the ones reachable at the Large Hadron Collider (LHC) and at new-generation hadron accelerators.
This permits us to access kinematic regions where $\sqrt{s} \gg m_{\cal Q}$. The limit in which $\sqrt{s} \gg \{Q \} \gg \Lambda_{\rm QCD}$\footnote{Here $s$ is the center-of-mass energy squared, $\{Q \}$ represents the (set of) perturbative scale(s) characterizing the process, and $\Lambda_{\rm QCD}$ stands for the QCD scale parameter.} is known as \emph{Regge--Gribov} or \emph{semi-hard} regime\tcite{Gribov:1983ivg,Celiberto:2017ius}.
As it is well known, in this regime large logarithms of the ratio $s/Q^2$ appear to all orders the perturbative series with a power increasing with the $\alpha_s$ order, thus spoiling its convergence. This simply means that the standard fixed-order approach for the computation of hard scattering cross sections is no longer a reliable framework to build theoretical predictions up. The need for a resummation to all orders that takes into account the effect of these logarithms, led authors of Ref.~\cite{Fadin:1975cb,Kuraev:1976ge,Kuraev:1977fs,Balitsky:1978ic} to develop the previously mentioned BFKL approach, which is now established as the the most powerful formalism for the description of the semi-hard QCD sector.
Its validity holds within the leading approximation (LLA), which means resumming all terms proportional to $\alpha_s^n \ln (s/Q^2)^n$, and within the next-to-leading approximation (NLA), which means including all terms proportional to $\alpha_s^{n+1} \ln (s/Q^2)^n$.

When the observed particles are widely separated in rapidity, cross sections for hadronic processes take a peculiar factorized form, given as the convolution of two impact factors, related to the fragmentation of each colliding particle to an identified final-state object, and a process-independent Green's function, which encodes the resummation of energy logarithms. The BFKL Green's function satisfies an integral evolution equation, whose kernel is known up to NLO for any fixed, not growing with $s$, momentum transfer $t$ and for any possible two-gluon color state in the $t$-channel\tcite{Fadin:1998py,Ciafaloni:1998gs,Fadin:1998jv,Fadin:2000kx,Fadin:2000hu,Fadin:2004zq,Fadin:2005zj}.
Impact factors instead depend on processes, so they representing the most challenging part of the calculation. So far, the number of impact factors calculated within NLO accuracy is quite small.

Evidence of the onset of the BFKL dynamics was provided in the analysis of differential distributions for the hadroproduction of two objects widely separated in rapidity. An incomplete list of them includes: the inclusive detection of two light jets featuring large transverse momenta and well separated in rapidity (Mueller--Navelet channel\tcite{Mueller:1986ey}), for which several phenomenological studies have appeared so far~(see, \emph{e.g.},~Refs.\tcite{Colferai:2010wu,Caporale:2012ih,Ducloue:2013hia,Ducloue:2013bva,Caporale:2013uva,Caporale:2014gpa,Colferai:2015zfa,Caporale:2015uva,Ducloue:2015jba,Celiberto:2015yba,Celiberto:2015mpa,Celiberto:2016ygs,Celiberto:2016vva,Caporale:2018qnm,Celiberto:2022gji}), the inclusive emission of a light di-hadron system\tcite{Celiberto:2016hae,Celiberto:2016zgb,Celiberto:2017ptm,Celiberto:2017uae,Celiberto:2017ydk}, multi-jet tags\tcite{Caporale:2015vya,Caporale:2015int,Caporale:2016soq,Caporale:2016vxt,Caporale:2016xku,Celiberto:2016vhn,Caporale:2016djm,Caporale:2016lnh,Caporale:2016zkc}, hadron-plus-jet\tcite{Bolognino:2018oth,Bolognino:2019cac,Bolognino:2019yqj,Celiberto:2022kxx}, Higgs-plus-jet\tcite{Celiberto:2020tmb,Celiberto:2021fjf,Celiberto:2021tky,Celiberto:2021txb,Celiberto:2020rxb,Celiberto:2021xpm,Celiberto:2022qbh}, heavy-light di-jet system\tcite{Bolognino:2021mrc,Bolognino:2021hxx}, Drell--Yan-plus-jet~\cite{Golec-Biernat:2018kem}, heavy-flavored hadrons\tcite{Celiberto:2017nyx,Bolognino:2019yls,Bolognino:2019ccd,Celiberto:2021dzy,Celiberto:2021fdp,Bolognino:2022wgl}, and notably $J/\psi$-plus-jet production\tcite{Boussarie:2017oae}.

A major problem rising in phenomenological studies of semi-hard processes via the high-energy resummation is connected to the fact that NLA corrections both to the BFKL Green's function and impact factors have almost the same size and opposite sign of pure LLA terms. This generates instabilities in the BFKL series unstable which become strongly manifest when renormalization and factorization scales are varied from their \emph{natural} values, namely the ones dictated by kinematics.
Although employing some scale-optimization method, such as the Brodsky--Lepage--Mackenzie (BLM) procedure~\cite{Brodsky:1996sg,Brodsky:1997sd,Brodsky:1998kn,Brodsky:2002ka} effectively helped to reduce these instabilities in reactions featuring the emissions of light jets and/or hadrons\tcite{Ducloue:2013bva,Caporale:2014gpa,Celiberto:2016hae,Celiberto:2017ptm}, it turned out that the given optimal values for scales were by far higher than the natural ones\tcite{Celiberto:2020wpk}. This led to a lowering of cross sections of one or more orders of magnitude, thus hampering any possibility of doing precision studies.

A first clue of a reached stability of semi-hard observables at natural scales was observed quite recently in reactions featuring the detection of particles with a large transverse mass, as Higgs bosons~\cite{Celiberto:2020tmb,Celiberto:2022zdg} (see Refs.\tcite{Hentschinski:2020tbi,Celiberto:2022fgx} for novel NLO calculations) and heavy-quark jets~\cite{Bolognino:2021mrc}. Due to the lack of NLO calculations for the corresponding impact factors, these analyses were performed with a partial NLA accuracy. A strong evidence of stabilization effects in full NLA calculation emerged in a recent study on $\Lambda_c$-baryon production~\cite{Celiberto:2021dzy}
In particular, it was pointed out that the peculiar pattern of VFNS FFs depicting the production of the charmed baryon acts as a fair stabilizer of the high-energy series.
This result was subsequently confirmed also for bottomed hadrons~\cite{Celiberto:2021fdp}.

With the aim of hunting for the aforementioned stabilizing effects in semi-hard emissions of heavy-quarkonium states, in this article we investigate the high-energy behavior of the inclusive hadroproduction of a $\JPsi$ or a $\Yps$ accompanied by a light-quark jet at the LHC.
The two detected objects are widely separated in rapidity.
We remark that only the \emph{direct} $\JPsi$ and $\Yps$ production components are considered in our study.
Both the meson and the jet feature large transverse momenta, and they are well separated in rapidity. 
Kinematic typical of current LHC analyses feature moderate values of partons' longitudinal-momentum fractions.\footnote{Although partons' longitudinal-momentum fractions can go down to around $10^{-4}$ for quarkonium and $10^{-5}$ for jet emissions, due to kinematics, numerical tests have shown that typical values giving the major contribution to cross sections rage from $10^{-3}$ to $0.5$, namely where both small-$x$ and large-$x$ effects are not relevant.} This justify a description in terms of collinear PDFs.
From a purely collinear-factorization viewpoint, the large required transverse momenta makes valid using a variable-flavor number-scheme (VFNS) approach\tcite{Mele:1990cw,Cacciari:1993mq}, in which the cross section for the production of a light parton is constructed and then convoluted with a FF that describes the transition from light quarks to heavy bound states ($\JPsi$ or $\Yps$ in our case). Then, making use of NRQCD, the quarkonium FF is constructed as a product between short-distance coefficient functions, which encode the resummation of DGLAP-type logarithms, and the corresponding LDMEs. We will employ a recent NLO determination for heavy-quark to $\JPsi$ and $\Yps$ FFs\tcite{Zheng:2019dfk}, together with the corresponding gluon to $\JPsi$ and $\Yps$ functions\tcite{Braaten:1993rw}.
From a high-energy perspective, large rapidity intervals in the final state lead to non-negligible exchanges of transverse momenta in the $t$-channel which call for a $p_T$-factorization procedure, naturally provided by the BFKL resummation within NLA accuracy.
Combining all these ingredient makes a full NLO treatment feasible. Our hybrid high-energy and collinear factorization, already set as the reference formalism for the description of inclusive two-particle semi-hard emissions, is thus enriched with a new element, the NRQCD, which provides us with a useful model for quarkonium fragmentation.
Another formalism, close in spirit with our hybrid factorization and suited for single forward detections, was proposed in Refs.\tcite{Deak:2009xt,vanHameren:2022mtk}.

As already mentioned, the fragmentation mechanism becomes increasingly competitive as well as the transverse momentum of tagged particle grows.
We remind, however, that since we are neglecting the heavy-quark mass at the level of the hard part, our formalism is not suited to properly describe the intermediate region, where $|{\vec p}_T| \sim m_Q$. Here powers of the $\sqrt{m_Q^2 + |{\vec p}_T|^2}/|{\vec p}_T|$ ratio need to be taken into account. In this sense, our study is complementary to the one proposed in Ref.\tcite{Boussarie:2017oae} (see also preliminary results in Refs.\tcite{Boussarie:2015jar,Boussarie:2016gaq}), where the inclusive semi-hard $\JPsi$-plus-jet production was investigated by making use of the short-distance $(Q \bar Q)$ mechanism for the quarkonium production.
In that work massive impact factors for the transitions $(g \to Q \bar{Q}  \to \JPsi)$ and $(g \to Q \bar{Q} g \to \JPsi + g)$, encoding the corresponding LDMEs, where computed and then used to consider both color-singlet and color-octet configurations of the meson.\footnote{Doubly off-shell LO impact factors for inclusive emissions of quarkonia in central regions of rapidity were calculated in Refs.\tcite{Hagler:2000dd,Kniehl:2006sk}.} In such a calculation all powers of $\sqrt{m_Q^2 + |{\vec p}_T|^2}/|{\vec p}_T|$ ratio are present, but since the final state features no DGLAP evolution, logarithms are not resummed. In the future, a possible matching between the two approaches could help us to enhance the description of heavy-flavor in the context of the high-energy resummation.

Coming back to our large-$p_T$ fragmentation treatment, possible ambiguities could emerge.
On the one side, a pure collinear-factorization description relies on the fragmentation of a single parton to the identified hadron. At large transverse momenta the parton is always treated as a light one and the hadronization mechanism is portrayed by zero-mass (ZM) VFNS FFs that evolve with DGLAP.
On the other side, a NRQCD FF always contains the perturbative splitting of the parton produced via the hard scattering to a massive $(Q \bar Q)$ pair.
Synthesizing these two viewpoints could hide some complications which are currently matter of investigation\tcite{JPsiFFs:2022}.
We leave these insights to future studies, and take the NRQCD functions of Ref.\tcite{Zheng:2019dfk} as proxy models for $\JPsi$ and $\Yps$ collinear FFs.

Furthermore, one might argue that our analysis is not complete due to the absence of the CO contribution.
Indeed, Ref.\tcite{Zheng:2019dfk} contains the FF of the constituent heavy quark, $Q$, to a color-singlet quarkonium state, $\cal Q$, namely the contribution proportional to the $^3S_1^{(1)}$ LDME (see Section\tref{ssec:FFs_JAlg} for more details), and the same holds for the gluon to quarkonium FF\tcite{Braaten:1993rw}.
We stress, however, that our primary goal is hunting for stabilizing effects of high-energy resummed distributions for $\JPsi$ and $\Yps$ production under higher-order corrections and energy-scale variations.
Therefore we rely on the CSM contribution only. The inclusion of higher Fock-state contributions in our single-parton fragmentation description is postponed to a forthcoming analysis.

The structure of this article reads as follows. In Section~\ref{sec:theory} we introduce our theoretical setup, presenting the highlights of our hybrid high-energy and collinear factorization (Section~\ref{ssec:cross_section}), as well as details on the heavy-meson fragmentation mechanism and the light-jet selection function (Section~\ref{ssec:FFs_JAlg}). In Section~\ref{sec:results}, we discuss our phenomenological study of differential distributions. Finally (Section~\ref{sec:conclusions}) we come out with a summary and future challenges.

\section{Theoretical setup}
\label{sec:theory}

\begin{figure*}[!t]
\centering
\includegraphics[width=0.65\textwidth]{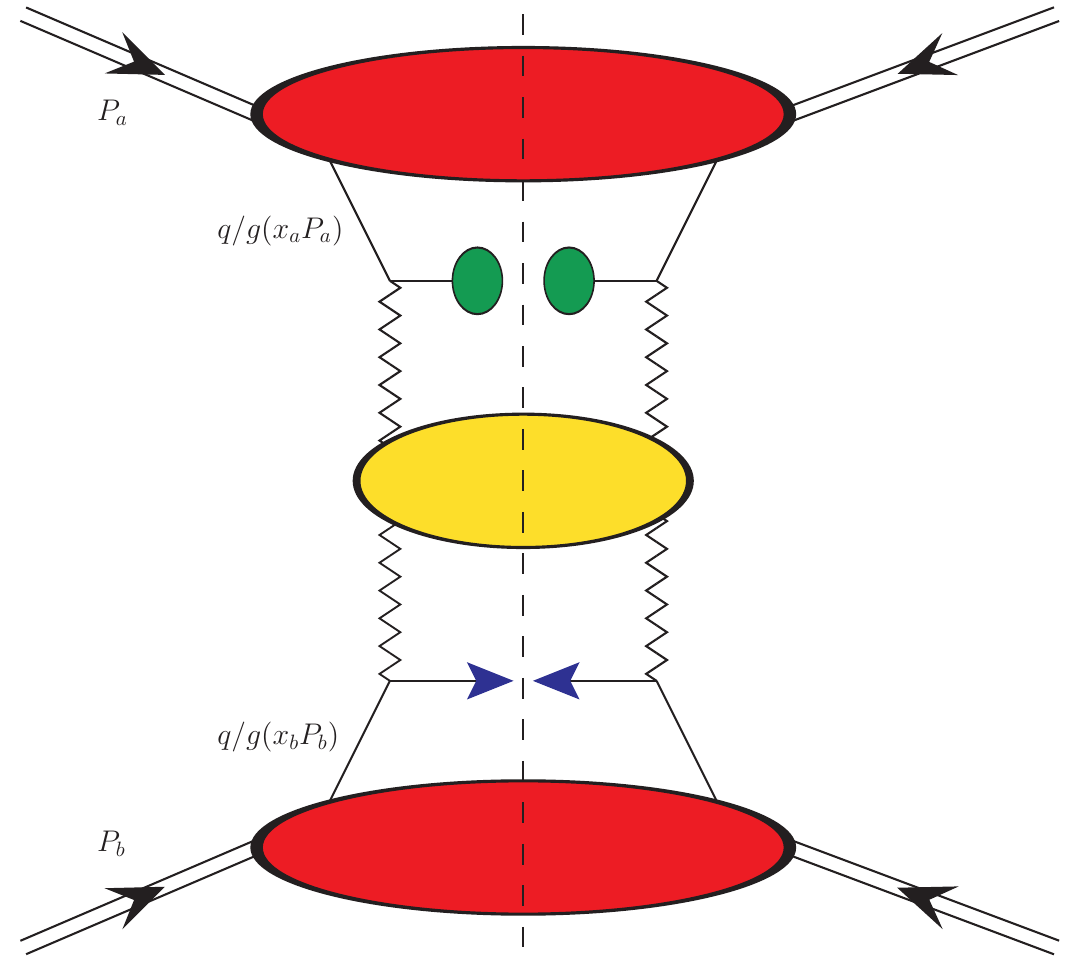}

\caption{Hybrid high-energy/collinear factorization for the inclusive quarkonium $+$ jet production. Red blobs refer to proton collinear PDFs, green blobs stand for quarkonium single-parton FFs, while blue arrows denote a light-flavored jet emission. The BFKL ladder, given by the yellow blob, is connected to impact factors via Reggeon (zigzag) lines. Diagrams were done by making use of the {\tt JaxoDraw 2.0} interface~\cite{Binosi:2008ig}.}
\label{fig:process}
\end{figure*}

In this Section we present theoretical ingredients to build our observables. First we give analytic expressions of azimuthal-angle coefficients for the considered reaction (Fig.\tref{fig:process}), calculated by means of our hybrid high-energy and collinear factorization~(Section\tref{ssec:cross_section}). Then we discuss our choice for quarkonium FFs and for the jet algorithm~(Section\tref{ssec:FFs_JAlg}).

\subsection{High-energy resummed cross section}
\label{ssec:cross_section}

The process under investigation is
\begin{equation}
\label{process}
    p(P_a) + p(P_b) \rightarrow \Q(p_\Q, y_\Q) + X + {\rm jet}(p_J, y_J) \; ,
\end{equation}
where $p(P_{a,b})$ stands for an initial proton with momentum $P_{a,b}$, $\Q(p_\Q, y_\Q)$ is a $\JPsi$ or a $\Yps$ emitted with momentum $p_\Q$ and rapidity $y_\Q$, the light jet is tagged with momentum $p_J$ and rapidity $y_J$, and $X$ denotes all the undetected products of the reaction. High observed transverse momenta, $|\vec p_{\Q,J}|$, together with a large rapidity separation, $\DY = y_\Q - y_J$, are required conditions to get a diffractive semi-hard configuration in the final state. Furthermore the transverse-momentum ranges need to be enough large to ensure the validity of description of the quarkonium production mechanism in terms of single-parton VFNS collinear fragmentation.  

The momenta of the two colliding protons are taken as Sudakov vectors satisfying $P_a^2= P_b^2=0$ and $2 (P_a\cdot P_b) = s$, and the four-momenta of final-state particles can be decomposed as
\begin{equation}\label{sudakov}
p_{\Q,J} = x_{\Q,J} P_{a,b} + \frac{\vec p_{\Q,J}^{\,2}}{x_{\Q,J} s}P_{b,a} + p_{\Q,J\perp} \ , \quad
p_{\Q,J\perp}^2=-\vec p_{\Q,J}^{\,2}\;,
\end{equation}
where the outgoing particle longitudinal momentum fractions, $x_{\Q,J}$, are connected to the corresponding rapidities via the relation
$y_{\Q,J}=\pm\frac{1}{2}\ln\frac{x_{\Q,J}^2 s}
{\vec p_{\Q,J}^2}$. One has $\drv y_{\Q,J} = \pm \frac{\drv x_{\Q,J}}{x_{\Q,J}}$, and $\DY \equiv y_\Q - y_J = \ln\frac{x_\Q x_J s}{|\vec p_\Q||\vec p_J|}$.

In a genuine QCD collinear treatment, the LO cross section of our process (Eq.\eref{process}) is cast as a convolution of the partonic hard subprocess with the parent-proton PDFs and the quarkonium FFs
\begin{equation}
\label{sigma_collinear}
\begin{split}
\frac{\drv\sigma^{\rm LO}_{\rm coll.}}{\drv x_\Q\drv x_J\drv ^2\vec p_\Q\drv ^2\vec p_J}
= \hspace{-0.25cm} \sum_{\alpha,\beta=q,{\bar q},g}\int_0^1 \hspace{-0.20cm} \drv x_a \int_0^1 \hspace{-0.20cm} \drv x_b\ f_\alpha\left(x_a\right) f_\beta\left(x_b\right)
\int_{x_\Q}^1 \hspace{-0.15cm} \frac{\drv z}{z}D^{\Q}_{\alpha}\left(\frac{x_\Q}{z}\right) 
\frac{\drv {\hat\sigma}_{\alpha,\beta}\left(\hat s\right)}
{\drv x_\Q\drv x_J\drv ^2\vec p_\Q\drv ^2\vec p_J}\;.
\end{split}
\end{equation}
Here the $\alpha,\beta$ indices indicate the parton species (quarks $q = u, d, s, c, b$; antiquarks $\bar q = \bar u, \bar d, \bar s, \bar c, \bar b$; or gluon $g$), $f_{\alpha,\beta}\left(x, \mu_F \right)$ are the colliding-proton PDFs and $D^{\Q}_{\alpha}\left(x/z, \mu_F \right)$ denote the outgoing-quarkonium FFs; $x_{a,b}$ are the longitudinal-momentum fractions of the partons initiating the hard subprocess and $z$ the longitudinal fraction of the single parton that fragments into $\Q$. Finally, $\drv\hat\sigma_{\alpha,\beta}\left(\hat s \right)$ stands for the partonic cross section, with $\hat s \equiv x_a x_b s$ the squared center-of-mass energy of the partonic collision.
For the sake of brevity, the explicit dependence of PDFs, FFs and partonic cross section on the factorization scale, $\mu_F$, has been dropped.

At variance with collinear factorization, in order to obtain the expression for the high-energy resummed cross section in our hybrid framework, one needs first to consider the high-energy factorization which is naturally encoded in the BFKL formalism, and then enhance the description by embodying collinear PDFs and FFs.
We start by writing the differential cross section as a Fourier sum of the so-called azimuthal-angle coefficients
\begin{equation}
 \label{dsigma_Fourier}
 \frac{\drv \sigma}{\drv y_\Q \drv y_J \drv \vec p_\Q \drv \vec p_J \drv \phi_\Q \drv \phi_J} =
 \frac{1}{(2\pi)^2} \left[{\cal C}_0 + 2 \sum_{n=1}^\infty \cos (n \varphi)\,
 {\cal C}_n \right]\, ,
\end{equation}
with $\varphi_{\Q,J}$ the azimuthal angles of the detected particles and $\varphi = \phi_\Q - \phi_J - \pi$.
The azimuthal coefficients ${\cal C}_n \equiv \CnNLA$ can be calculated in the BFKL approach and they encode the resummation of energy logarithms up to the NLA accuracy. A NLA formula obtained in the $\MSb$ renormalization scheme reads (for details on the derivation see, \emph{e.g.}, Ref.~\cite{Caporale:2012ih})
\[
 \CnNLA = \int_0^{2\pi} \drv \phi_\Q \int_0^{2\pi} \drv \phi_J\,
 \cos (n \varphi) \,
 \frac{\drv \sigma_{\rm NLA}}{\drv y_\Q \drv y_J\, \drv |\vec p_\Q| \, \drv |\vec p_J| \drv \phi_\Q \drv \phi_J}\;
\]
\[
 = \frac{e^{\DY}}{s} 
 \int_{-\infty}^{+\infty} \drv \nu \, e^{{\DY} \bar \alpha_s(\mu_R)\left\{\chi(n,\nu)+\bar\alpha_s(\mu_R)
 \left[\bar\chi(n,\nu)+\frac{\beta_0}{8 N_c}\chi(n,\nu)\left[-\chi(n,\nu)+\frac{10}{3}+4\ln\left(\frac{\mu_R}{\sqrt{|\vec p_\Q| |\vec p_J|}}\right)\right]\right]\right\}}
\]
\begin{equation}
\label{Cn_NLA_MSb}
 \times \, \alpha_s^2(\mu_R) \, 
 \left[
 c_\Q^{\rm NLO}(n,\nu,|\vec p_\Q|, x_1)[c_J^{\rm NLO}(n,\nu,|\vec p_J|,x_2)]^*\,
 + \bar \alpha_s^2(\mu_R)
 \, \DY
 \frac{\beta_0}{4 N_c}\chi(n,\nu)f(\nu)
 \right] \;,
\end{equation}
with $\bar \alpha_s(\mu_R) \equiv \alpha_s(\mu_R) N_c/\pi$, $N_c$ the color number and $\beta_0 = 11N_c/3 - 2 n_f/3$ the first coefficient of the QCD $\beta$-function.
A two-loop running-coupling setup with $\alpha_s\left(M_Z\right)=0.11707$ and a dynamic flavor number $n_f$ is employed.
The expression for the LO BFKL characteristic function reads
\begin{equation}
\chi\left(n,\nu\right)=2\left\{\psi\left(1\right)-{\rm Re} \left[\psi\left( i\nu+\frac{1}{2}+\frac{n}{2} \right)\right] \right\}
\label{chi}
\end{equation}
where $\psi(z) = \Gamma^\prime(z)/\Gamma(z)$ is the logarithmic derivative of the Gamma function. The $\bar\chi(n,\nu)$ function in Eq.\eref{Cn_NLA_MSb} contains NLO corrections to the BFKL kernel and was calculated in Ref.~\cite{Kotikov:2000pm} (see also Ref.~\cite{Kotikov:2002ab}).
The two functions
\begin{equation}
\label{IFs}
c_{\Q,J}^{\rm NLO}(n,\nu,|\vec p\,|,x) =
c_{\Q,J}(n,\nu,|\vec p\,|,x) +
\alpha_s(\mu_R) \, \hat c_{\Q,J}(n,\nu,|\vec p\,|,x)
\end{equation}
are the impact factors for the production of a heavy-quarkonium state and for the emission of a light jet.
Their LO parts read
\[
c_\Q(n,\nu,|\vec p\,|,x) 
= 2 \sqrt{\frac{C_F}{C_A}}
(|\vec p\,|^2)^{i\nu-1/2}\,\int_{x}^1\frac{\drv \zeta}{\zeta}
\left( \frac{\zeta}{x} \right)
^{2 i\nu-1} 
\]
\begin{equation}
\label{LOQIF}
 \times \left[\frac{C_A}{C_F}f_g(\zeta)D_g^\Q\left(\frac{x}{\zeta}\right)
 +\sum_{\alpha=q,\bar q}f_\alpha(\zeta)D_\alpha^\Q\left(\frac{x}{\zeta}\right)\right] 
\end{equation}
and
\begin{equation}
 \label{LOJIF}
 c_J(n,\nu,|\vec p\,|,x) =  2 \sqrt{\frac{C_F}{C_A}}
 (|\vec p\,|^2)^{i\nu-1/2}\,\left(\frac{C_A}{C_F}f_g(x)
 +\sum_{\beta=q,\bar q}f_\beta(x)\right) \;,
\end{equation}
respectively. The $f(\nu)$ function is defined in terms of the logarithmic derivative of LO impact factors
\begin{equation}
 f(\nu) = \frac{i}{2} \, \frac{\drv}{\drv \nu} \ln\left(\frac{c_\Q}{c_J^*}\right) + \ln\left(|\vec p_\Q| |\vec p_J|\right) \;.
\label{fnu}
\end{equation}
The remaining functions in Eq.~(\ref{Cn_NLA_MSb}) are the NLO impact-factor corrections, $\hat c_{\Q,J}$.
The NLO correction to the $\Q$ impact factor is calculated in the light-quark limit\tcite{Ivanov:2012iv}. This option is fully consistent with our VFNS treatment, provided that the $p_\Q$ values at work are much larger than the heavy-quark mass (charm or bottom, see Section\tref{sec:results} for more details).
Our choice for the jet NLO impact factor is discussed at the end of Section\tref{ssec:FFs_JAlg}.

We present for completeness the expression for the azimuthal coefficients in the pure LLA approximation, namely by neglecting NLO terms in the BFKL kernel and impact factors
\begin{equation}
\label{Cn_LLA_MSb}
  \CnLLA = \frac{e^{\DY}}{s} 
 \int_{-\infty}^{+\infty} \drv \nu \, e^{{\DY} \bar \alpha_s(\mu_R)\chi(n,\nu)} \, \alpha_s^2(\mu_R) \, c_\Q(n,\nu,|\vec p_\Q|, x_1)[c_J(n,\nu,|\vec p_J|,x_2)]^* \,.
\end{equation}

In Eqs.~(\ref{Cn_NLA_MSb})-(\ref{Cn_LLA_MSb}) it is highlighted the way how our hybrid factorization is realized. The cross section is factorized \emph{à la} BFKL as a convolution between the Green's function and impact factors, the latter ones encoding the collinear functions.

We use expressions given in this Section at the \emph{natural} scales provided by the process. In particular we set $\mu_F = \mu_R = \mu_N \equiv \sqrt{m_{\Q \perp} |\vec p_J|}$, with $m_{\Q \perp} = \sqrt{ m_\Q^2 + |\vec p_\Q|^2}$ the quarkonium transverse mass. We fix $m_\Q \equiv m_{\JPsi} = 3.0969$ GeV or $m_\Q \equiv m_{\Yps} = 9.4603$ GeV, according to the emitted meson.
Collinear PDFs are obtained via the {\tt MMHT14} NLO parameterization~\cite{Harland-Lang:2014zoa} as provided by the {\tt LHAPDFv6.2.1} library~\cite{Buckley:2014ana}.
All calculations of our azimuthal coefficients are done in the $\MSb$ renormalization scheme.

As a supplementary analysis, we will present values of energy scales obtained by applying the BLM optimization method\tcite{Brodsky:1996sg,Brodsky:1997sd,Brodsky:1998kn,Brodsky:2002ka}. We will not calculate our observables at these scales, but we will compare these last with the ones typical of inclusive emissions of other heavy-flavored hadrons, thus corroborating the statement that heavy-flavor collinear FFs act as stabilizers of the high-energy resummation.

The BLM method prescribes that the \emph{optimal} renormalization-scale value, $\mu_R^{\rm BLM}$, is the one that allows us to remove the $\beta_0$ dependence of a given observable.
In Ref.\tcite{Caporale:2015uva} a general procedure to apply the BLM prescription on BFKL-resummed azimuthal coefficients was set up. 
The condition for the BLM scale setting on a given coefficient, $C_n$, is the solution of the following integral equation
\begin{equation}
\label{Cn_beta0_int}
  C_n^{(\beta)}(s, \DY) = 
  \int \drv \Phi(y_{\Q,J}, |\vec p_{\Q,J}|, \DY) \,
  \, {\cal C}_n^{(\beta)}  = 0 \, ,
\end{equation}
where $\drv \Phi(y_{\Q,J}, |\vec p_{\Q,J}|, \DY)$ is the final-state differential phase space (see Section\tref{sec:results}),
\[
 {\cal C}^{(\beta)}_n
 \propto \!\!
 \int^{\infty}_{-\infty} \!\!\drv\nu\,e^{\DY \bar \alpha^{\rm MOM}_s(\mu^{\rm BLM}_R)\chi(n,\nu)}
 c_\Q(n,\nu,|\vec p_\Q|,x_1)[c_J(n,\nu,|\vec p_J|,x_2)]^*
\]
\begin{equation}
\label{Cn_beta0}
 \times \, \left[{\omega}(\nu) + \bar \alpha^{\rm MOM}_s(\mu^{\rm BLM}_R) \DY \: \chi(n,\nu) \left(- \frac{\chi(n,\nu)}{4} + {\omega}(\nu) \right) \right] \, ,
\end{equation}
and
\begin{equation}
\label{upsilon_nu}
{\omega}(\nu) = \frac{f(\nu)}{2} - \frac{1}{6} - \frac{2}{3} I + \ln \left( \frac{\mu^{\rm BLM}_R}{\sqrt{|\vec p_\Q| |\vec p_J|}} \right) \, .
\end{equation}
In Eq.\eref{Cn_beta0} $\bar \alpha_s^{\rm{MOM}} = N_c/\pi \, \alpha_s^{\rm{MOM}}$ stands for the expression of the strong coupling in the momentum renormalization scheme (MOM), given by inverting the relation
\begin{equation}
\label{as_MOM}
 \as^{\MSb} = \as^{\rm MOM} \left[ 1 +  (\tau^{\beta} + \tau^{\rm conf}) \frac{\as^{\rm MOM}}{\pi} \right] \;,
\end{equation}
with
\begin{equation}
\label{T_bc}
\tau^\beta = - \left( \frac{1}{2} + \frac{I}{3} \right) \beta_0
\end{equation}
and
\begin{equation}
\tau^{\rm conf}= \frac{N_c}{8}\left[ \frac{17}{2}I +\frac{3}{2}\left(I-1\right)\xi
+\left( 1-\frac{1}{3}I\right)\xi^2-\frac{1}{6}\xi^3 \right] \; ,
\end{equation}
where $I=-2\int_0^1\drv z \frac{\ln~z}{z^2-z+1} \simeq 2.3439$ and $\xi$ is gauge parameter fixed at zero in the following.
We write the optimal renormalization scale as $\mu_R^{\rm BLM} = C_\mu^{\rm BLM} \mu_N$, with $\mu_N$ the process natural scale defined previously, and we look for values of the  $C_\mu^{\rm BLM}$ parameter which solve Eq.~(\ref{Cn_beta0_int}). 

\subsection{Quarkonium fragmentation and jet selection function}
\label{ssec:FFs_JAlg}

We build our NLO collinear FF sets for the \emph{direct} $\JPsi$ or $\Yps$ meson production by taking, as a starting point, the recent work done in Ref.\tcite{Zheng:2019dfk}. There, a NLO calculation was performed for the heavy-quark FF depicting the transition $c \to \JPsi$ or the $b \to \Yps$ one, where $c$ ($b$) indistinctly refer to the charm (bottom) quark and its antiquark. It essentially relies on the NRQCD factorization formalism taken with NLO accuracy (see, \emph{e.g.}, Refs.\tcite{Thacker:1990bm,Bodwin:1994jh,QuarkoniumWorkingGroup:2004kpm,Brambilla:2010cs,Pineda:2011dg,Brambilla:2020ojz} and references therein), which allows us to write the FF function of a parton $i$ fragmenting into a heavy quarkonium $\cal Q$ with longitudinal fraction $z$ as\footnote{Within the same formalism, collinear FFs for $\bar b$ and $c$ quarks fragmenting into $B_c$ and $B_c^*$ mesons were calculated with LO\tcite{Chang:1992bb,Braaten:1993jn,Ma:1994zt} and NLO accuracy\tcite{Zheng:2019gnb} (see Ref.\tcite{Celiberto:2022keu} for a recent application at the High-Luminosity LHC)}.
\begin{equation}
 \label{FF_NRQCD}
 D^{\cal Q}_i(z, \mu_F) = \sum_{[n]} {\cal D}^{\cal Q}_{i}(z, \mu_F, [n]) \langle {\cal O}^{\cal Q}([n]) \rangle \;.
\end{equation}
In Eq.\eref{FF_NRQCD}, ${\cal D}_{i}(z, \mu_F, [n])$ denotes the perturbative short-distance coefficient containing terms proportional to $\ln (\mu_F/m_{\cal Q})$ (to be resummed via DGLAP evolution), $\langle {\cal O}^{\cal Q}([n]) \rangle$ stands for the non-perturbative NRQCD LDME, and $[n] \equiv \,^{2S+1}L_J^{(c)}$ represents the quarkonium quantum numbers in the spectroscopic notation (see, \emph{e.g.}, Ref.\tcite{Bugge:1986xw}), the $(c)$ superscript identifying the color state, singlet (1) or octet (8).
Limiting ourselves to a spin-triplet (vector) and color-singlet quarkonium state, $^3S_1^{(1)}$, the analytic form of the initial-scale FF depicting the constituent heavy-quark to quarkonium transition, $Q \to {\cal Q}$ (we inclusively refer to $c \to \JPsi$ or $b \to \Yps$), reads (for details on its derivation, see Sections~II~and~III of Ref.\tcite{Zheng:2019dfk})
\begin{equation}
 \label{FF_Q-to-onium}
 D^{\cal Q}_Q(z, \mu_F \equiv \mu_0) 
 = D^{\cal Q, {\rm LO}}_Q(z)
 + \frac{\alpha_s^3(3m_Q)}{m_Q^3} \, |{\cal R}_{\cal Q}(0)|^2 \, \Gamma_Q^{\cal Q, {\rm NLO}}(z) \;,
\end{equation}
with $m_c = 1.5$~GeV or~$m_b = 4.9$~GeV, and the NRQCD radial wave-function at the origin of the quarkonium state set to~$|{\cal R}_{\JPsi}(0)|^2 = 0.810$~GeV$^3$ or to~$|{\cal R}_{\Yps}(0)|^2 = 6.477$~GeV$^3$, according to potential-model calculations (Ref.\tcite{Eichten:1994gt} and references therein).
The expression for the LO initial-scale FF was originally calculated in Ref.\tcite{Braaten:1993mp} and reads
\begin{equation}
 \label{FF_Q-to-onium_LO}
 D^{\cal Q, {\rm LO}}_Q(z) = 
 \frac{\alpha_s^2(3m_Q)}{m_Q^3} \, \frac{8z(1-z)^2}{27\pi (2-z)^6} \, |{\cal R}_{\cal Q}(0)|^2 \, (5z^4 - 32z^3 + 72z^2 -32z + 16) \;,
\end{equation}
and the polynomial function $\Gamma_Q^{\cal Q, {\rm NLO}}(z)$ entering the expression for the NLO-FF correction is
\begin{eqnarray}
 \label{FF_Gamma_JPsi}
 \Gamma_Q^{\JPsi, {\rm NLO}}(z) 
 &=& - 9.01726z^{10} + 18.22777z^9 + 16.11858z^8 - 82.54936z^7 
 \nonumber \\
 &+& 106.57565z^6 - 72.30107z^5 + 28.85798z^4 - 6.70607z^3
 \nonumber \\
 &+& 0.84950z^2 - 0.05376z - 0.00205
\end{eqnarray}
or
\begin{eqnarray}
 \label{FF_Gamma_Yps}
 \Gamma_Q^{\Yps, {\rm NLO}}(z) 
 &=& - 14.00334z^{10} + 46.94869z^9 - 55.23509z^8 + 16.69070z^7 
 \nonumber \\
 &+& 22.09895z^6 - 26.85003z^5 + 13.41858z^4 - 3.50293z^3
 \nonumber \\
 &+& 0.46758z^2 - 0.03099z - 0.00226 \:.
\end{eqnarray}
Coefficients of $z$-powers in Eqs.\eref{FF_Gamma_JPsi}~and\eref{FF_Gamma_Yps} are obtained via a polynomial fit to the numerically-calculated NLO FFs.
Starting from $\mu_F \equiv \mu_0 = 3 m_Q$, in Ref.\tcite{Zheng:2019dfk} a DGLAP-evolved formula for the $D^{\cal Q}_Q(z, \mu_F)$ function was derived and then applied to phenomenological studies of $\JPsi$ and $\Yps$ production via $e^+ e^-$ single inclusive annihilation (SIA).

As pointed out in Ref.\tcite{Braaten:1994xb}, both $(c \to \JPsi)$ and $(g \to \JPsi)$ fragmentation channels are similar in size. The relative weight of the heavy-quark and gluon contributions is also driven by the size of the hard scattering producing these partons. Therefore, the number of large $p_T$-gluons emitted could be of the same order, if not larger, than the heavy-quark one.
Moreover, in a hadroproduction process such as the one considered in our study, the gluon FF is enhanced by the collinear convolution at LO with the corresponding gluon PDF (see Eq.\eref{LOQIF}). Thus we expect, in our case, a stronger sensitivity on the gluon-fragmentation channel with respect to the case of a SIA-like reaction.
Therefore, we include in our analysis also the contribution coming from the gluon fragmentation. The gluon to vector-quarkonium LO fragmentation mechanism start at $\alpha_s^3$, namely at the same order of the heavy-quark FF in Eq.\eref{FF_Q-to-onium}. The $(g \to {^3S_1^{(1)}} g g)$ fragmentation function was computed in Ref.\tcite{Braaten:1993rw} and reads
\begin{equation*}
    D_g^{\cal Q} (z, 2 m_Q) = \frac{5}{36 (2\pi)^2} \alpha_s^3(2 m_Q) \frac{|{\cal R}_{\cal Q}(0)|^2}{m_{\cal Q}^3} \int_0^z \drv \xi \int_{(\xi+z^2)/2z}^{(1+\xi)/2} \drv \tau \frac{1}{(1-\tau)^2 (\tau-\xi)^2 (\tau^2-\xi)^2} 
\end{equation*}
\begin{equation}
    \sum_{i=1}^{2} z^i \left[ f^{(g)}_i (\xi, \tau) + g^{(g)}_i (\xi, \tau) \frac{1+\xi-2 \tau}{2 (\tau-\xi) \sqrt{\tau^2-\xi}} \ln \left( \frac{\tau - \xi + \sqrt{\tau^2-\xi}}{\tau - \xi - \sqrt{\tau^2-\xi}} \right) \right] \; ,
\label{FF_g-to-onium}
\end{equation}
with the six $f^{(g)}_i$ and $g^{(g)}_i$ functions being given in Eqs.~(4)-(9) of Ref.\tcite{Braaten:1995cj}. We stress that the mechanism considered here is the \emph{direct} production of the quarkonium from the parent gluon. Another contribution to $\JPsi$-meson production in high energy processes, not considered in our analysis, is the production of a $P$-wave charmonium state $\chi_c$, followed by its radiative decay $\chi_c \to \JPsi + \gamma$ (see Refs.\tcite{Braaten:1993mp,Yuan:1994hn}). The different initial energy scales at which quarks and gluons FF are taken is due to the production mechanism itself. The heavy-quark fragmentation involves at least three heavy quarks in the final state (Fig.\tref{fig:FF_diagrams}, left panel), thus the running coupling in Eqs.\eref{FF_Q-to-onium} and\eref{FF_Q-to-onium_LO} is calculated at $\mu_R = 3m_Q$. Conversely, the gluon fragmentation involves two heavy quarks only (Fig.\tref{fig:FF_diagrams}, right panel), and this explains why the running coupling in Eq.\eref{FF_g-to-onium} is taken at $\mu_R = 2m_Q$.

For the present study we follow a twofold strategy.
As a first step, starting from the initial-scale FF in Eq.\eref{FF_Q-to-onium}, we generate the corresponding functions for all parton species. This is a required step in order to perform analyses by means of our high-energy VFNS treatment (see Section\tref{ssec:cross_section}). For a given quarkonium, $\JPsi$ or $\Yps$, we set the corresponding constituent heavy (anti-)quark FF, $D^{\JPsi}_c \equiv D^{\JPsi}_{\bar c}$ or $D^{\Yps}_b \equiv D^{\Yps}_{\bar b}$, to be equal to the parameterization given in Eq.\eref{FF_Q-to-onium} at the initial scale $\mu_0 = 3 m_Q$. Then, we compute the DGLAP-evolved functions via the {\tt APFEL++} library\tcite{Bertone:2013vaa,Carrazza:2014gfa,Bertone:2017gds}, thus getting for each meson a {\tt LHAPDF} set of FFs that embodies all parton flavors. From now, according to names of Authors of Ref.\tcite{Zheng:2019dfk}, we will refer to these sets as the $\JPsi$ and $\Yps$ {\tt ZCW19} NLO FF parameterizations. This first strategy of generating the remnant parton FFs only via DGLAP evolution could be justified by the fact that, by definition, the lowest Fock state of a quarkonium is a $(Q \bar Q)$ pair, whose contribution heavily dominates among the other flavors. Moreover, this choice is in line with general assumptions made in the extraction of FFs of other heavy-flavored hadrons, as $D^{(*)}$~mesons\tcite{Kniehl:2005de,Kneesch:2007ey,Anderle:2017cgl,Soleymaninia:2017xhc}, $B$~mesons\tcite{Kniehl:2007erq,Kniehl:2011bk} and $\Lambda_c$\tcite{Kniehl:2020szu}~baryons.

As a second step, having in mind the outcome of the previously mentioned study on gluon fragmentation to vector quarkonium\tcite{Braaten:1994xb}, we improve our FF treatment by starting the DGLAP evolution from the gluon input of Eq.\eref{FF_g-to-onium} at the initial scale $\mu_F = 2 m_Q$. Then, when the energy grows up to reach $\mu_F = 3 m_Q$, the heavy-quark channel is opened and the corresponding FF (Eq.\eref{FF_Q-to-onium}) starts its evolution. In this way, for each meson we get another {\tt LHAPDF} FF set, named {\tt ZCW19$^+$}, which embodies both the gluon and the heavy-quark NRQCD inputs to fragmentation.
For our phenomenology we will mainly employ the {\tt ZCW19$^+$} functions, and we will use the {\tt ZCW19} ones for cross-checks.

\begin{figure*}[!t]

   \includegraphics[scale=0.53,clip]{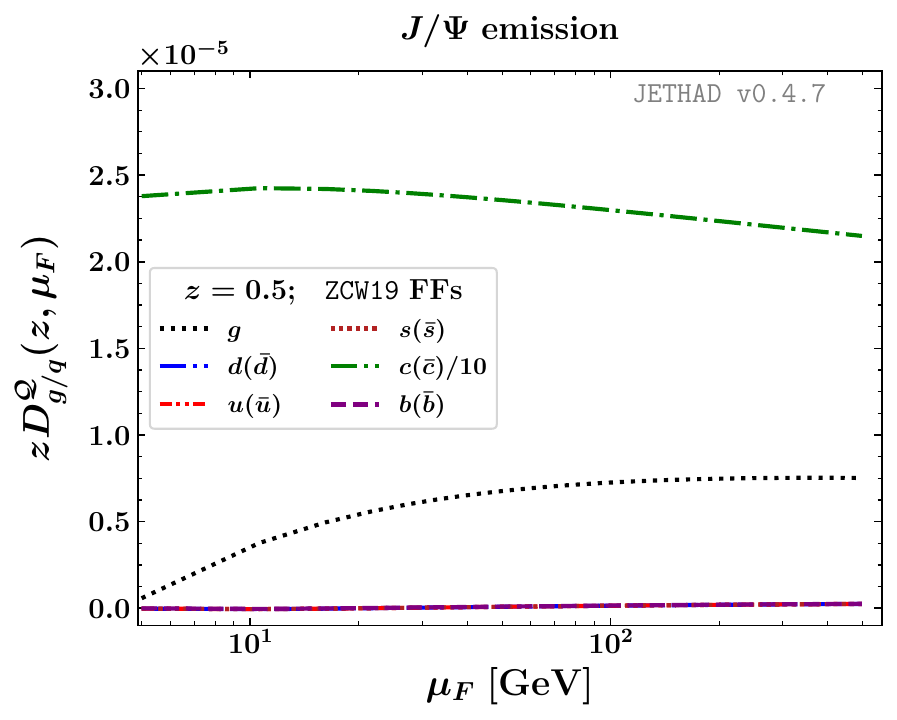}
   \includegraphics[scale=0.53,clip]{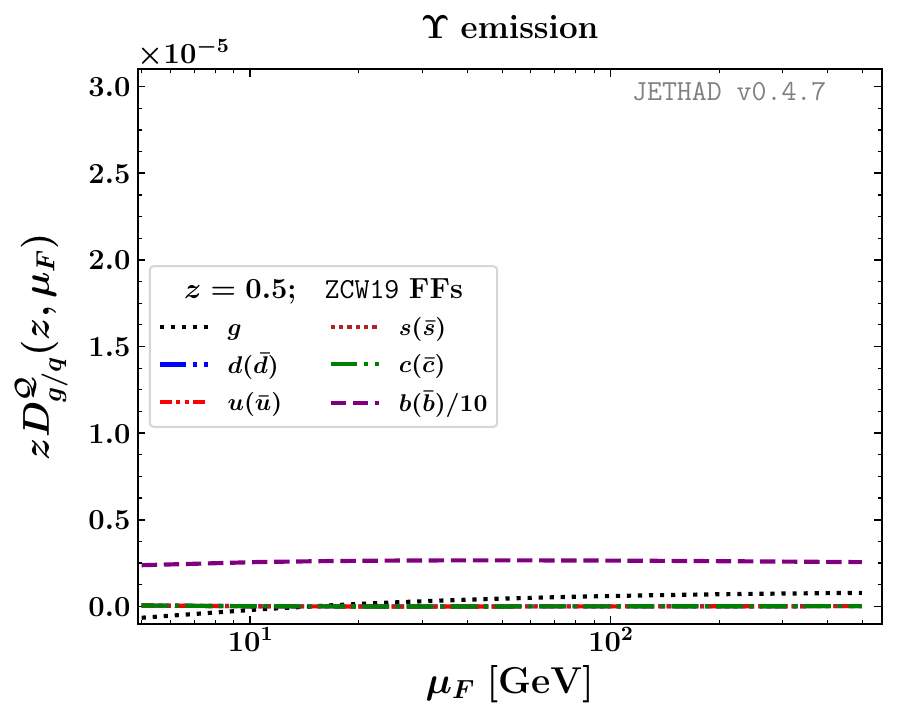}

   \includegraphics[scale=0.53,clip]{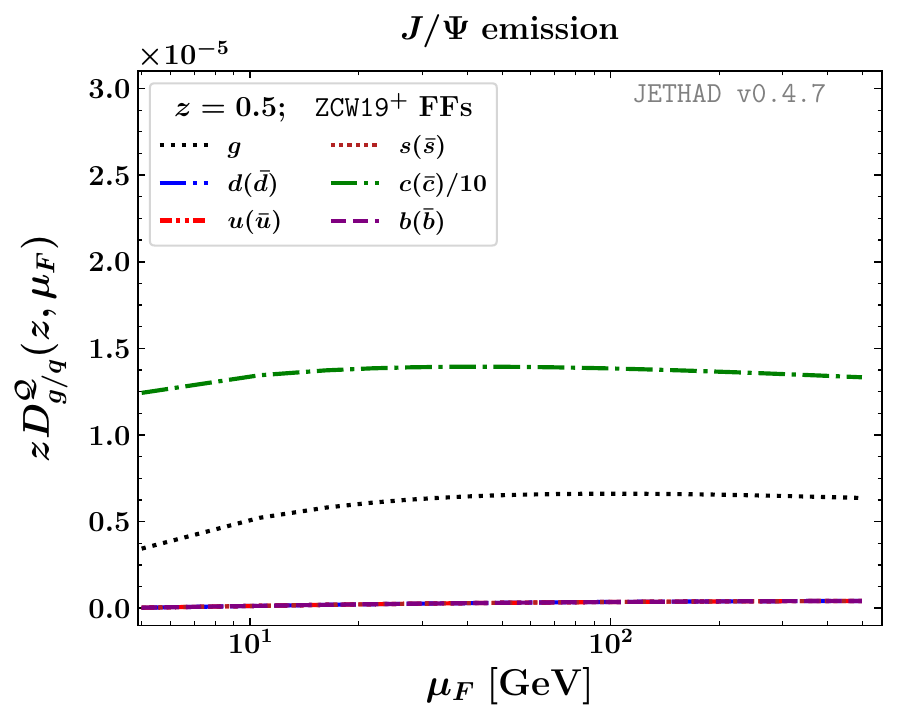}
   \includegraphics[scale=0.53,clip]{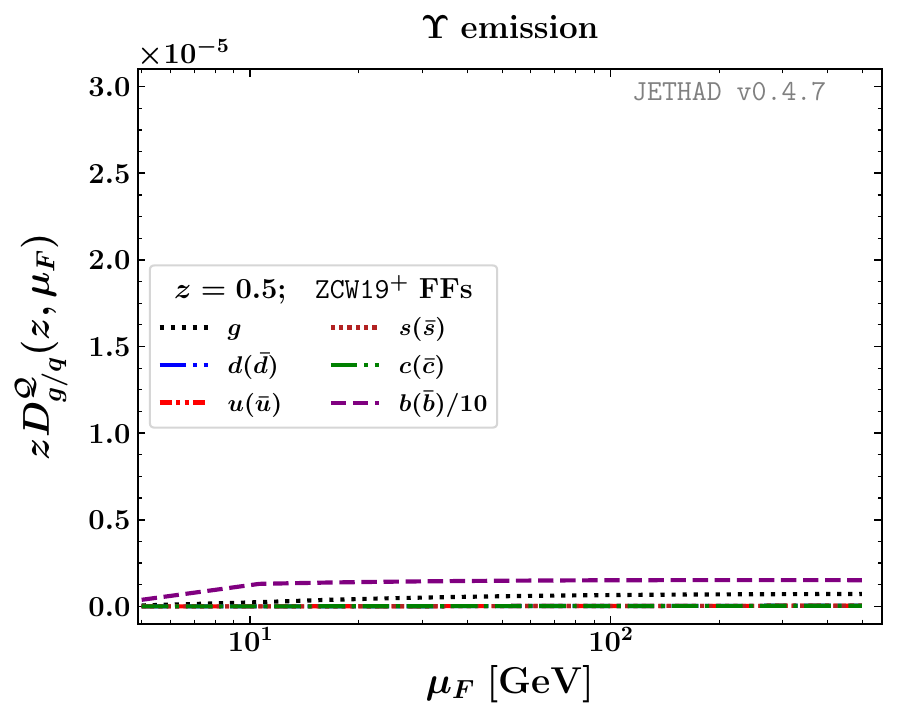}

\caption{Energy-scale dependence of {\tt ZCW19} (upper) and {\tt ZCW19$^+$} (lower)} NLO FFs depicting $\JPsi$ (left) and $\Yps$ (right) emission, for $z = 5 \times 10^{-1}$.
\label{fig:FFs_psv}
\end{figure*}

For comparison with previous studies on inclusive semi-hard emissions of heavy-flavored hadrons\tcite{Celiberto:2021dzy,Celiberto:2021fdp}, in Fig.\tref{fig:FFs_psv} we plot $\mu_F$-dependence of our {\tt ZCW19} (upper panels) and {\tt ZCW19$^+$} (lower panels) FF sets for a value of the quarkonium longitudinal momentum fraction that roughly corresponds to the average value of $z$ at which collinear FFs are typically probed in the consider kinematic configurations, namely $z = 5 \times 10^{-1} \simeq \langle z \rangle$.
As expected, the FF for the constituent heavy (anti-)quark, $c(\bar c)$ for $\JPsi$ (left panels) and $b(\bar b)$ for $\Yps$ (right panels), strongly prevails over the other ones. Notably, as already observed in previous analyses on $\Lambda_c$ baryons\tcite{Celiberto:2021dzy} and bottomed hadrons\tcite{Celiberto:2021fdp}, the gluon FF grows with $\mu_F$ up to reach a plateau. Although being much smaller than the constituent heavy-quark one, the gluon-FF contribution is strongly enhanced by the gluon PDF in the diagonal convolution embodied in the LO quarkonium impact factor~(see Eq.~(\ref{LOQIF})) and this feature is kept also at NLO, where $(qg)$ and $(gq)$ non-diagonal channels are active, but their weight cannot compensate the $(gg)$ one.
As pointed out in the aforementioned studies (see Section~3.4 of Ref.\tcite{Celiberto:2021dzy} and Appendix~A of Ref.\tcite{Celiberto:2021fdp}), the $\mu_F$-behavior of the gluon FF plays a crucial role in the stabilization of the high-energy resummation under higher-order corrections. More in particular, a clear evidence was brought that a smooth-behaved, non-decreasing with $\mu_F$ gluon FF leads to a strong stabilizing effect on rapidity-differential cross sections, that also reflects in a partial stabilization final-state azimuthal distributions. Furthermore, this translates in a weaker sensitivity of semi-hard observables on renormalization- and factorization-scale variations.
In view of these consideration, we expect that these stabilizing effects could fairly emerge in our quarkonium-plus-jet phenomenological analysis.
We observe that the inclusion of the initial-scale gluon FF leads to a global lowering of the DGLAP-evolved heavy-quark one, which is smaller in the {\tt ZCW19$^+$} case. The evolved gluon FF is qualitatively similar in both case, the corresponding {\tt ZCW19$^+$} function being larger (smaller) than the {\tt ZCW19} one in the low (high) $\mu_F$ range.

We describe jet emissions at NLO perturbative accuracy in terms of a reconstruction algorithm calculated within the the so-called ``small-cone'' approximation (SCA)~\cite{Furman:1981kf,Aversa:1988vb}, \emph{i.e.} for a small-jet cone aperture in the rapidity/azimuthal angle plane. More in particular, we adopt the version derived in Ref.~\cite{Ivanov:2012ms}, which is \emph{infrared-safe} up to NLO perturbative and well-suited for numerical computations, with a cone-jet function selection~\cite{Colferai:2015zfa} and for the jet-cone radius fixed at ${\cal R}_J = 0.5$, as usually done in recent experimental analyses at CMS\tcite{Khachatryan:2016udy}. The expression for the NLO jet vertex can be obtained, \emph{e.g.}, by combining Eq.~(36) of Ref.\tcite{Caporale:2012ih} with Eqs.~(4.19)-(4.20) of Ref.\tcite{Colferai:2015zfa}.

\begin{figure}[!t]

   \includegraphics[scale=0.53,clip]{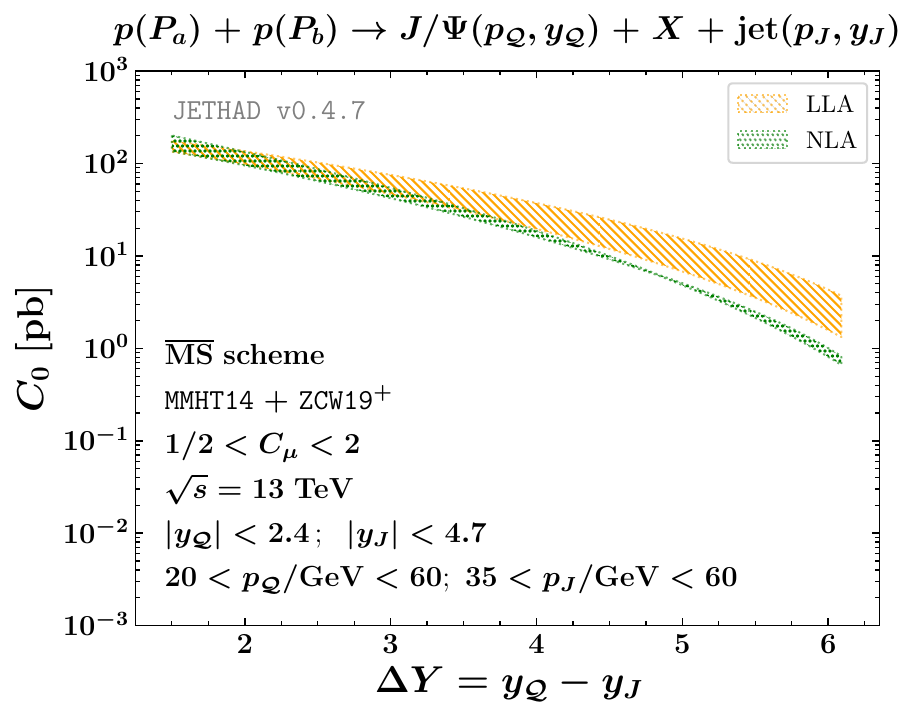}
   \hspace{0.225cm}
   \includegraphics[scale=0.53,clip]{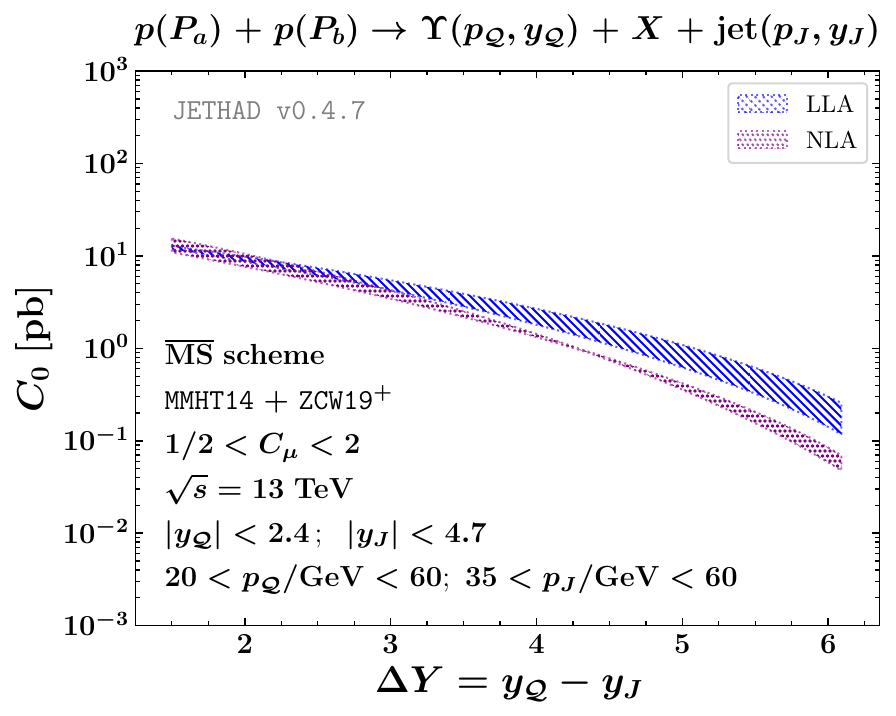}

   \includegraphics[scale=0.53,clip]{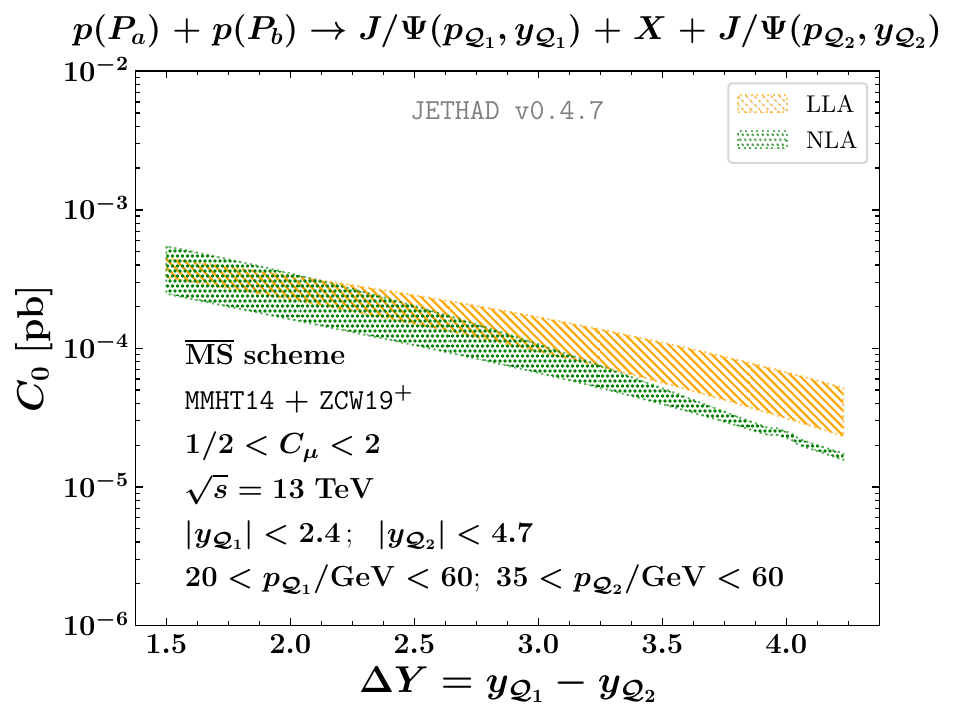}
   \includegraphics[scale=0.53,clip]{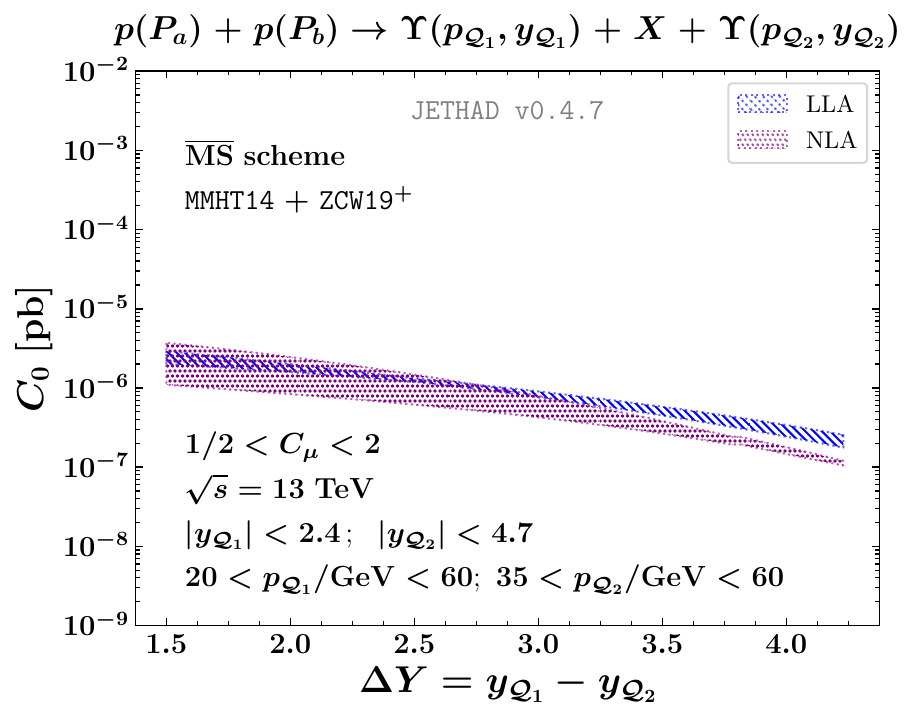}

   \includegraphics[scale=0.53,clip]{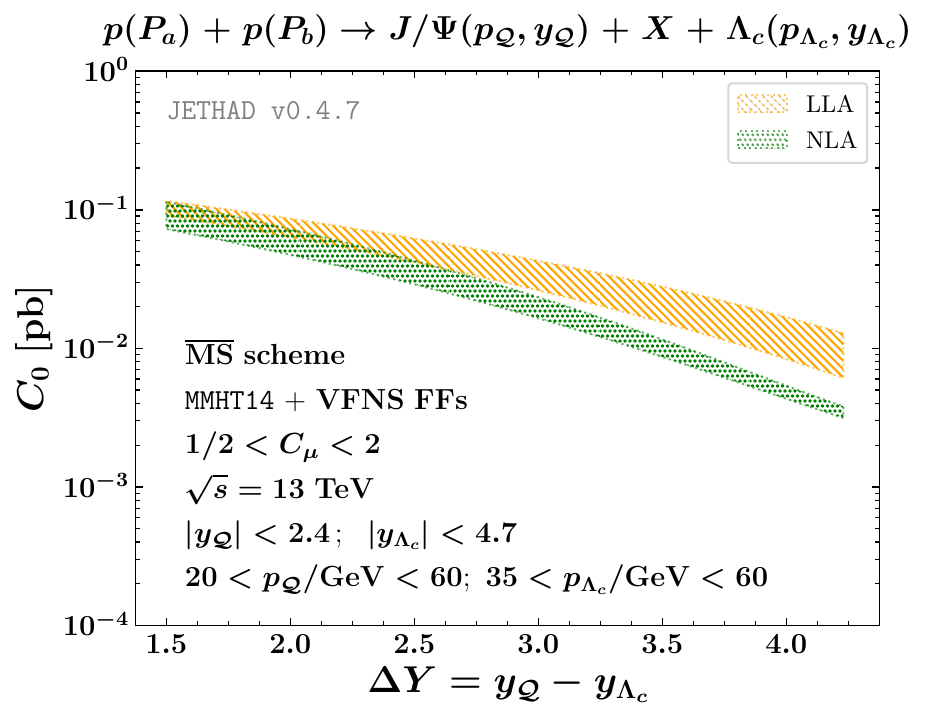}
   \hspace{0.14cm}
   \includegraphics[scale=0.53,clip]{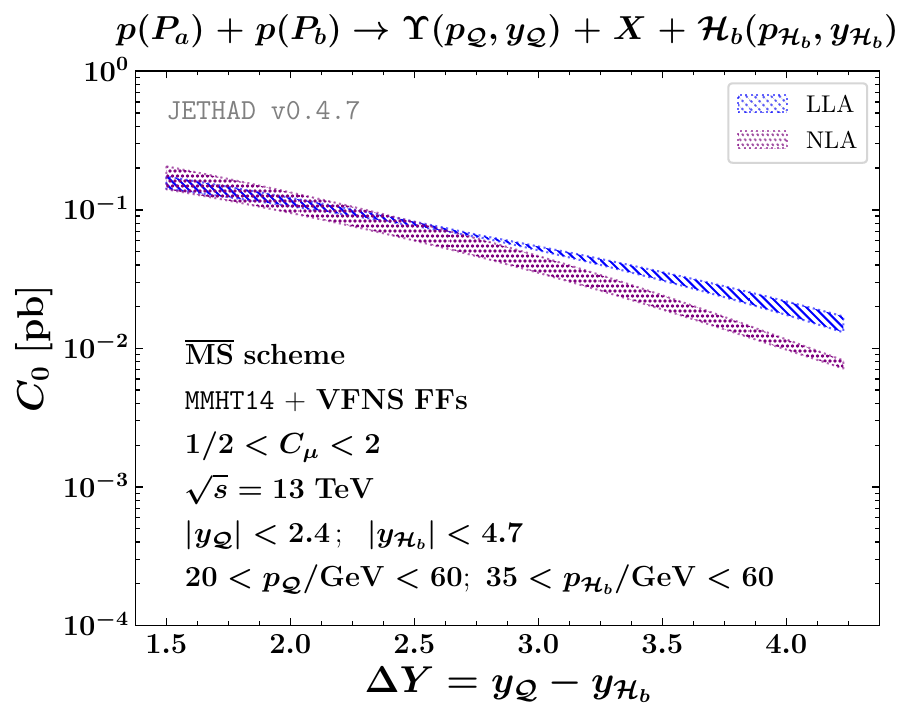}

\caption{Upper panels: $\DY$-distribution in the $\JPsi$~$+$~jet (left) and in the $\Yps$~$+$~jet (right) channel, for $\sqrt{s} = 13$ TeV.
Quarkonium fragmentation is described in terms of {\tt ZCW19$^+$} functions.
Central panels: the same observable in double quarkonium production channels.
Lower panels: predictions for the quarkonium~$+$~heavy-hadron detection.
Text boxes inside panels show transverse-momentum and rapidity ranges. The combined effect of scale variation and phase-space numerical integration is embodied in uncertainty bands.}
\label{fig:C0_HSA}
\end{figure}

\section{Numerical analysis}
\label{sec:results}

The numerical elaboration of all the considered observables was done by making use of the {\tt JETHAD} modular work package\tcite{Celiberto:2020wpk}.
The sensitivity of our results on scale variation was assessed by letting $\mu_R$ and $\mu_F$ to be around their \emph{natural} values or their BLM \emph{optimal} ones, up to a factor ranging from 1/2 to two.
The $C_{\mu}$ parameter entering plots represents the ratio $C_\mu = \mu_{R,F}/\mu_N$. Error bands in our figures embody the combined effect of scale variation and phase-space multi-dimensional integration, the latter being steadily kept below 1\% by the {\tt JETHAD} integrators.
All calculations of our observables were done in the $\MSb$ scheme. BLM scales are calculated by solving the integral equation\eref{Cn_beta0_int} in the MOM scheme.

\subsection{$\DY$-distribution}
\label{ssec:C0}

The first observable that we consider in our phenomenological analysis is the so-called $\DY$-distribution. It essentially corresponds to the $\varphi$-summed cross section differential in the rapidity interval, whose expression is obtained by integrating the ${\cal C}_0$ azimuthal coefficient (see Eq.\eref{Cn_NLA_MSb}) over rapidities and transverse momenta of the two emitted objects, and imposing the delta condition coming from fixing the value of $\DY$. One has
\begin{equation}
 \label{DY_distribution}
 C_0 =
 \int_{y_\Q^{\rm min}}^{y_\Q^{\rm max}} \hspace{-0.15cm} \drv y_\Q
 \int_{y_J^{\rm min}}^{y_J^{\rm max}} \hspace{-0.15cm} \drv y_J
 \int_{p_\Q^{\rm min}}^{p_\Q^{\rm max}} \hspace{-0.15cm} \drv |\vec p_\Q|
 \int_{p_J^{\rm min}}^{p_J^{\rm max}} \hspace{-0.15cm} \drv |\vec p_J|
 \, \,
 \delta (\DY - (y_\Q - y_J))
 \, \,
 {\cal C}_0\left(|\vec p_\Q|, |\vec p_J|, y_\Q, y_J \right)
 \, .
\end{equation}
Following the choice made in previous works on forward emissions of heavy-flavored hadrons\tcite{Celiberto:2021dzy,Celiberto:2021fdp}, we set the rapidity range of the quarkonium so that its detection is done only by the CMS barrel detector and not by endcaps, \emph{i.e.}~$|y_{\cal Q}| < 2.4$.
Then, for the sake of consistency with the VFNS treatment, where energy scales need to be much larger than thresholds for DGLAP evolution of heavy quarks, we allow the quarkonium transverse momentum to be in the range 20~GeV~$< |\vec p_{\cal Q}| <$~60~GeV.
The light jet is tagged in kinematic configurations typical of current studies at the CMS detector\tcite{Khachatryan:2016udy}, namely~35~GeV~$< p_J <$~60~GeV and $|y_J| < 4.7$.

In upper panels of Fig.\tref{fig:C0_HSA} we compare the NLA $\DY$-behavior of $C_0$ with the corresponding prediction at LLA.
Predictions were obtained by making use of the {\tt ZCW19$^+$} set. We note that values of $C_0$ are everywhere larger than 0.5 pb in the $J/\psi$-plus-jet channel (left) and $5\times10^{-2}$ pb in the $\Upsilon$-plus-jet one (right). This leads to a very promising statistics, although being substantially lower than the one for heavy-baryon and heavy-light meson emissions\tcite{Celiberto:2021dzy,Celiberto:2021fdp}.
The downtrend with $\DY$ of our distributions both at LLA and NLA is a common feature of all the hadronic semi-hard reactions investigated so far. It emerges as the interplay of two competing effects. On the one hand, the pure high-energy evolution leads to the well-known growth with energy of partonic cross sections. On the other hand, collinear parton distributions and fragmentation functions quench hadronic cross sections when $\DY$ increases.
We observe a partial stabilization of the high-energy series, with NLA bands almost overlapped to LLA ones at lower values of $\DY$, and their mutual distance becoming wider in the large-$\DY$ range. This feature is in line with previous studies on semi-hard heavy-flavor production, where the impressive stability of $C_0$ on next-to-leading order corrections observed in di-hadron channels ($\Lambda_c + \Lambda_c$\tcite{Celiberto:2021dzy} and double $b$-flavored hadron\tcite{Celiberto:2021fdp}) is partially lost when light-jet emissions are instead considered.
When a vector quarkonium is produced, the uncertainty band of $C_0$ at NLA is almost always smaller than the LLA one. 
To further examine this aspect, we compare our quarkonium-plus-jet predictions with results obtained when the light jet is replaced by another heavy-flavored hadron produced in the same final-state kinematic window. 
In central panels of Fig.\tref{fig:C0_HSA} we show the $\DY$-behavior of $C_0$ for a double quarkonium production (double $\JPsi$ or $\Yps$).
Here, the behavior of $C_0$ is similar to the one observed in quarkonium-plus-jet channel.
In particular, LLA and NLA bands decouple when $\DY$ grows.
In lower panels of Fig.\tref{fig:C0_HSA} we consider the production of a quarkonium plus a hadron sharing the same heavy flavor.
More in particular, the $\JPsi$ is accompanied by a $\Lambda_c^\pm$ baryon, described in terms of the {\tt KKSS19} NLO FF parameterization\tcite{Kniehl:2020szu}, while the $\Upsilon$ is emitted together a $b$-flavored hadron (${\cal H}_b$), portrayed by the {\tt KKSS07} NLO FF set\tcite{Kniehl:2011bk,Kramer:2018vde}.
We observe that, at large $\DY$, the NLA band for the $\JPsi$-plus-$\Lambda_c$ production is slightly wider than the double $\JPsi$ one (left central panel) and has almost the same size of the $\JPsi$-plus-jet one (left upper panel).
Conversely, the NLA band for the $\Yps$-plus-${\cal H}_b$ tag is almost always smaller than the double $\Yps$ one (right central panel) and the $\Yps$-plus-jet one (right upper panel).
All these observations support the statement that the high-energy stabilizing power of $\JPsi$ collinear FFs is stronger than the $\Lambda_c$ FF one, while $\Yps$ FFs act as weaker stabilizers with respect to the ${\cal H}_b$ ones.

\begin{figure*}[!t]
\centering

   \includegraphics[scale=0.44,clip]{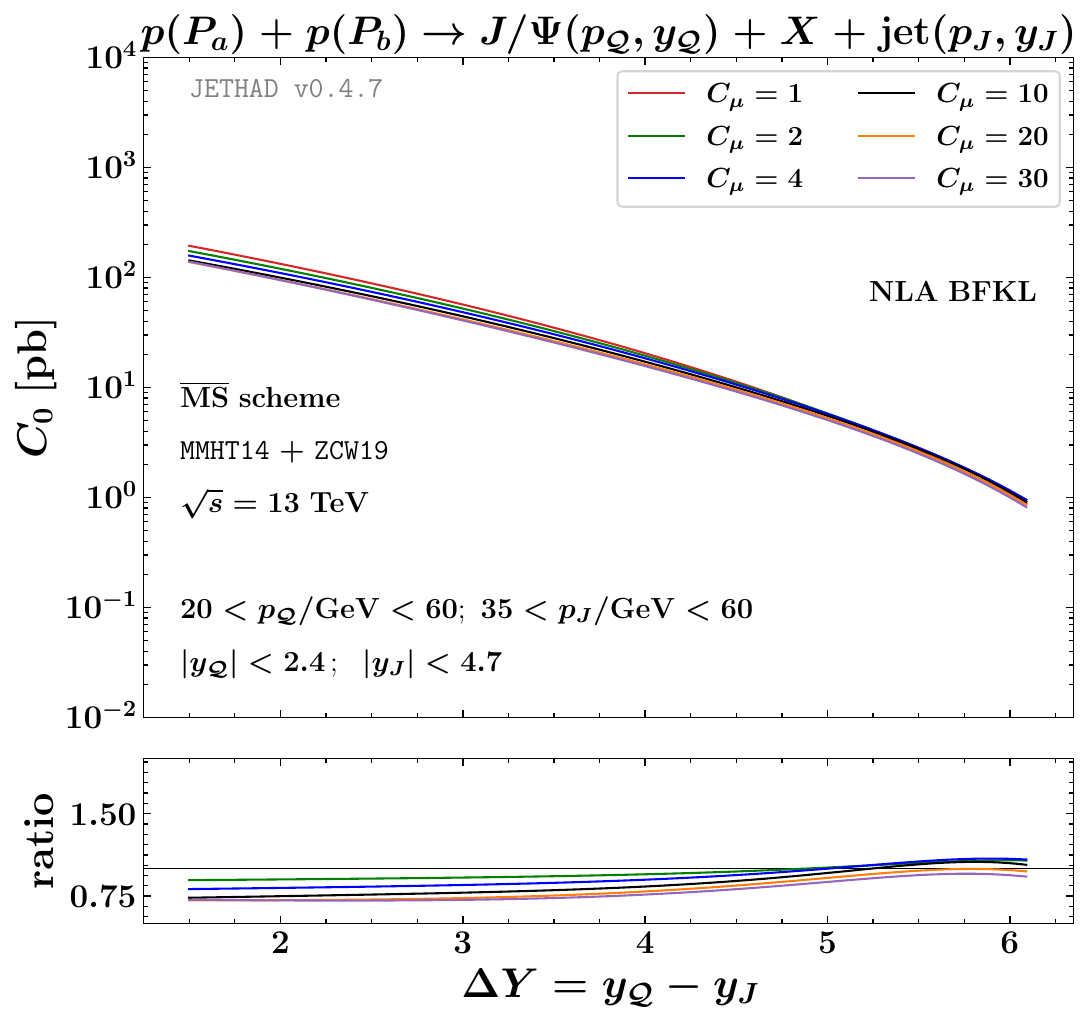}
   \includegraphics[scale=0.44,clip]{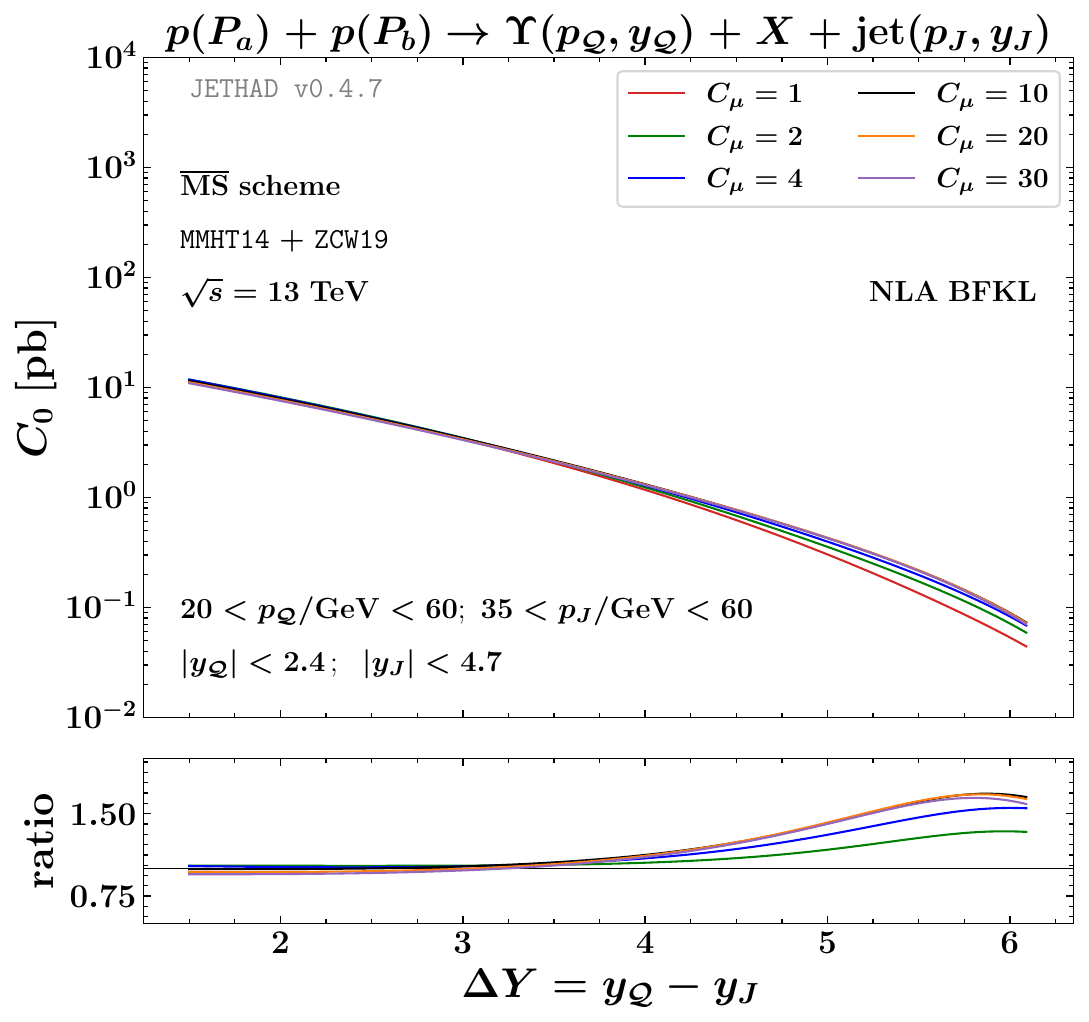}

   \includegraphics[scale=0.44,clip]{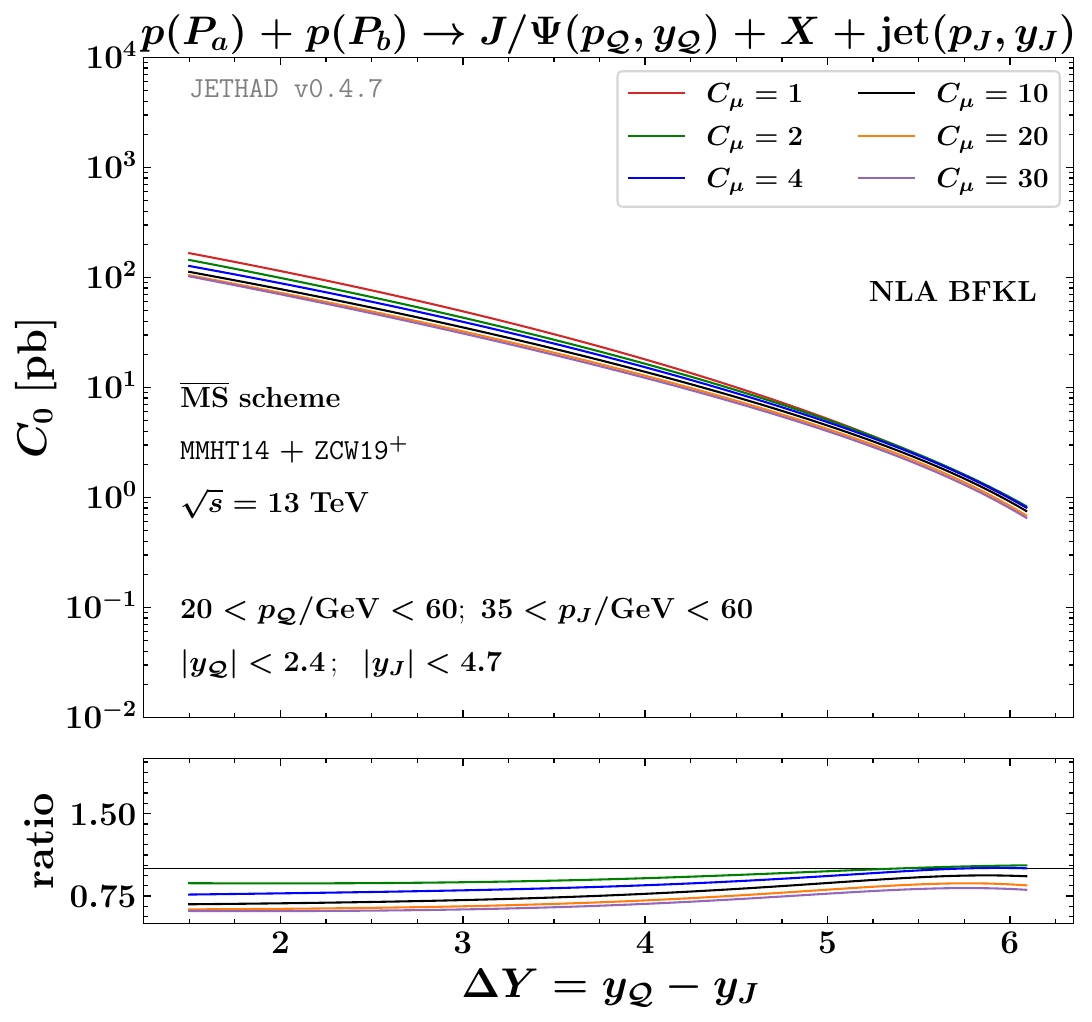}
   \includegraphics[scale=0.44,clip]{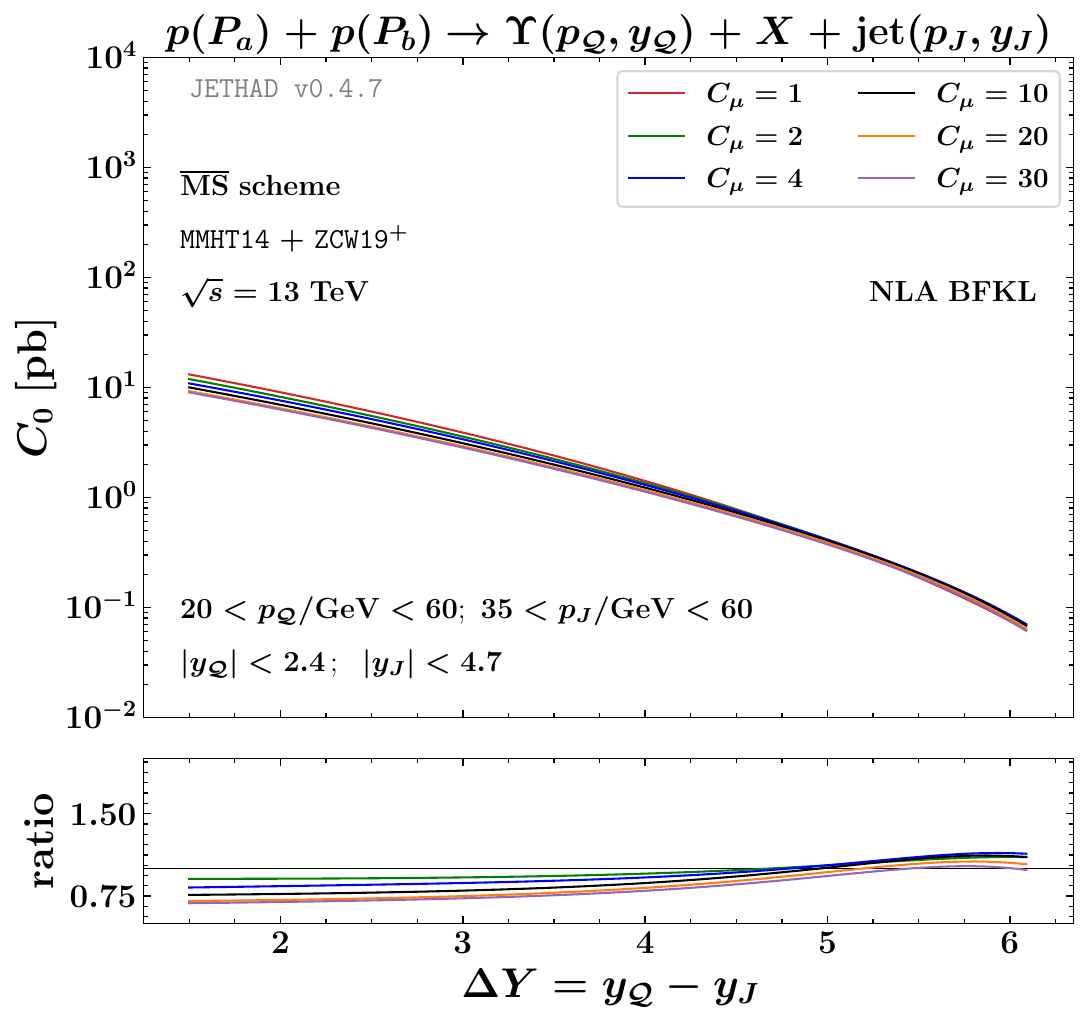}

\caption{$\DY$-distribution in the $\JPsi$~$+$~jet (left) and in the $\Yps$~$+$~jet (right) channel, for $\sqrt{s} = 13$ TeV.
Quarkonium fragmentation is described in terms of {\tt ZCW19} (upper) or {\tt ZCW19$^+$} (lower) functions. 
A study on progressive energy-scale variation in the range $1 < C_{\mu} < 30$ is made.
Ancillary panels below primary plots exhibit reduced distributions, namely divided by $C_0$ taken at $C_\mu = 1$.}
\label{fig:C0_psv}
\end{figure*}

\begin{figure*}[!t]
\centering

   \includegraphics[scale=0.53,clip]{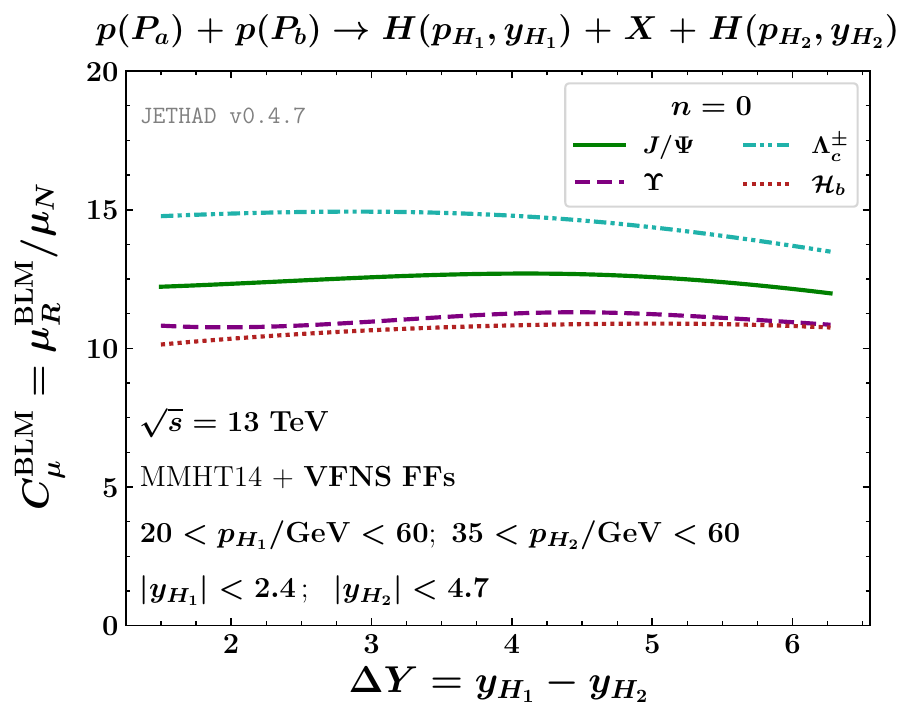}
   \includegraphics[scale=0.53,clip]{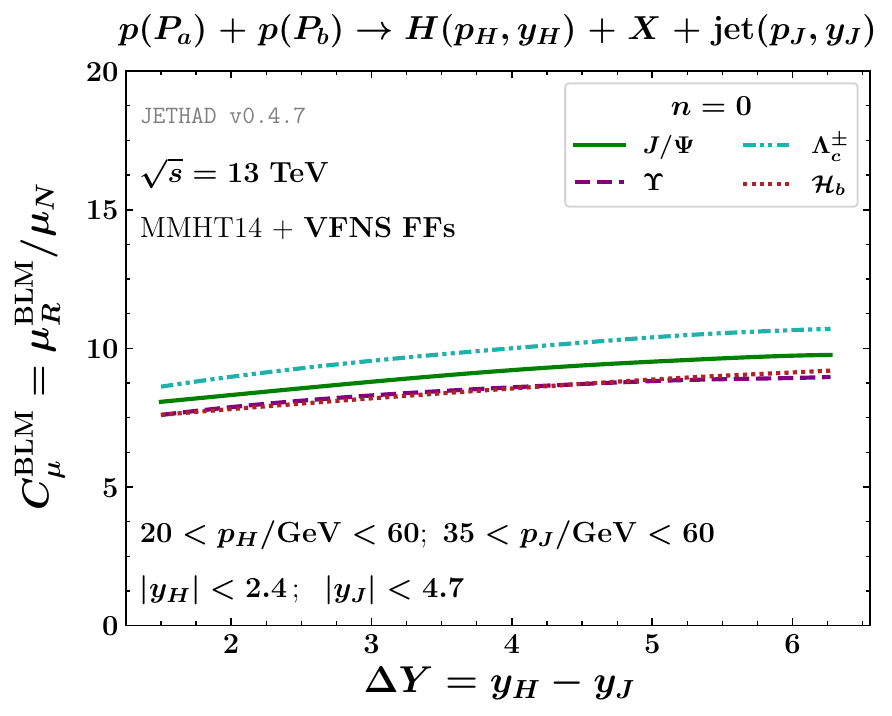}

\caption{BLM scales for the quarkonium-plus-jet (left) and the double quarkonium (right) production as functions of the rapidity separation, $\DY$, for $C_0$, and for $\sqrt{s} = 13$ TeV. Predictions for $\JPsi$ and $\Yps$ emissions, described in terms of {\tt ZCW19$^+$} FFs, are compared with configurations where other heavy-flavored hadron species are detected: $\Lambda_c$ baryons and ${\cal H}_b$ hadrons. Text boxes inside panels show transverse-momentum and rapidity ranges.}
\label{fig:BLM_scales_HSA}
\end{figure*}

In Fig.\tref{fig:C0_psv} we study the $\DY$-behavior of the NLA-resummed $\DY$-distribution under a progressive variation of $\mu_R$ and $\mu_F$ scales in the wider range given by $1 < C_\mu < 30$.
Upper (lower) plots contain predictions obtained via {\tt ZCW19$^{(+)}$} FFs. Ancillary panels below primary plots show the reduced distributions, \emph{i.e.} divided by $C_0$ taken at $C_\mu = 1$. As a first remark, we note that the magnitude of $C_0$ is almost independent on the choice of the quarkonium FF set. Notably, the sensitivity of {\tt ZCW19} predictions on the $C_\mu$ parameter is rather weak for the $J/\psi$-plus-jet channel, while it becomes more evident at larger values of $\DY$ for the $\Upsilon$-plus-jet reaction. Conversely, when {\tt ZCW19$^{+}$} functions are used, $J/\psi$-plus-jet emissions exhibit almost the same patter, while $\Upsilon$-plus-jet detections become less sensitive to scale variation when $\DY$ grows. This is an indication that employing a complete NRQCD input, built in terms of both initial-scale heavy-quark and gluon functions, leads to a stronger stabilizing pattern than the one achievable by relying on the quark input only.
Combining this information with the arguments presented above, we can easily explain the size of error bands for NLA results in upper panels of Fig.\tref{fig:C0_HSA}. Indeed, the major contribution to scale uncertainty comes, at least for the $J/\psi$-plus-jet emission, from variation of $\mu_R$ and $\mu_F$ in the $1/2 < C_\mu < 1$ lower sub-range. Here, the value of energy scales comes closer, even if still larger, to DGLAP-evolution thresholds connected to heavy-quark masses and this generates some instabilities which in turn lead to an increased scale-variation sensitivity of our predictions. 

As a supplementary analysis, we present results for the scale parameter $C_\mu \equiv C_\mu^{\rm BLM}$ obtained by applying the BLM optimization method on $C_0$. In Fig.\tref{fig:BLM_scales_HSA} we compare the $\DY$-dependence of $C_\mu^{\rm BLM}$ for our quarkonium production channels, described by means of the {\tt ZCW19$^+$} set, with the corresponding one for the emission of other $c$- and $b$-flavored non-quarkonium bound states, again $\Lambda_c$ baryons and $b$-hadrons.
In Refs.\tcite{Celiberto:2021dzy,Celiberto:2021fdp} it was shown that the BLM-scale values obtained when those heavy-flavored hadrons are detected in the final state are sensibly lower than the ones typical of lighter-hadron tags (see, \emph{e.g.}, Refs.\tcite{Bolognino:2018oth,Celiberto:2020rxb,Celiberto:2020wpk}).
Since the use of the BLM prescription operationally leads to a growth of $C_\mu^{\rm BLM}$ to dampen the weight of next-to-leading order corrections, processes where smaller BLM scales are found natively embody an intrinsic, partial stabilization of the high-energy series, before adopting optimization procedure.
The inspection of results presented in Fig.\tref{fig:BLM_scales_HSA} fairly confirms the statement that those stabilization effects are present also when $J/\psi$ and $\Upsilon$ mesons are considered.
More in particular, left panel of Fig.\tref{fig:BLM_scales_HSA} show how $C_0$-related BLM scales are clearly smaller in the double $J/\psi$ channel with respect to the double $\Lambda_c$ one, while they are slightly larger in the double $\Upsilon$ channel with respect to the double $b$-hadron one. Then, in line with studies done in Ref.\tcite{Celiberto:2021fdp}, $b$-flavored quarkonium and non-quarkonium detections provide us with smaller scales than the ones obtained for $c$-flavored tags. The same hierarchy is found in the heavy-hadron plus light jet channel (Fig.\tref{fig:BLM_scales_HSA}, right panel), matter of investigation in this article.

The general outcome of results presented in this Sections is that studies on forward $J/\psi$ and $\Upsilon$ emissions via {\tt ZCW19$^{(+)}$} NLO collinear FFs lead to a favorable statistics for the $\DY$-differential cross section. Effects of stabilization of the high-energy resummation under higher-order corrections and scale variation are presents and are even stronger, at least for the $c$-flavor channel, when quarkonia are detected instead of other heavy-flavored hadrons. At the same time, the simultaneous tag of a quarkonium together a light-flavored jet could partially shadow these effects, and further, dedicated studies on the improvement of the theoretical description of NLO jet emissions are needed.

Another relevant point that deserves a discussion is the weight of MPI contributions to cross sections at large $\DY$.
Let us focus on the double parton scattering (DPS), namely where at most two hard-scattering subprocesses are considered.
In Ref.\tcite{Ducloue:2015jba} it was shown that DPS contributions to Mueller--Navelet jets can be potentially relevant for $\DY$-distributions at large $s$ and low $p_T$, while, in the case of angular correlations between the two jets, they lead to results compatible with a NLA description based on single parton scattering only.
Analyses on double $\JPsi$\tcite{Lansberg:2014swa,Lansberg:2020rft}, $\JPsi$~$+$~$\Yps$\tcite{Lansberg:2020rft}, $\JPsi$~$+$~$Z$\tcite{Lansberg:2016rcx}, and $\JPsi$~$+$~$W$\tcite{Lansberg:2017chq} production processes have provided a clear indication that DPS effects are relevant and they need to be accounted for in studying the low-$p_T$ spectrum. Possible MPI signatures are expected also in triple $\JPsi$ events\tcite{dEnterria:2016ids,Shao:2019qob}.
Therefore, the search for DPS imprints in our vector-quarkonium plus jet observables is an important task that needs to be addressed in future studies.

\subsection{Azimuthal distribution}
\label{ssec:phi}

The study of the azimuthal-angle distribution in semi-hard two-particle final-state reactions is widely recognized as one of the most fertile grounds where to hunt for distinctive signals of the BFKL dynamics.
Here, the weight of undetected gluons strongly ordered in rapidity, accounted for the resummation of high-energy logarithms, becomes more and more relevant when the rapidity interval $\DY$ between the two objects grows. Thus, the decorrelation of the detected particles on the azimuthal plane increases at large $\DY$ and the number of back-to-back events diminishes.

First observables where the azimuthal-angle decorrelation was observed are the so called azimuthal-correlation moments, namely the ratios of azimuthal coefficients $R_{n0} \equiv C_n/C_0$. Here, $C_0$ is the $\varphi$-summed $\DY$-distribution introduced in Section\tref{ssec:C0}, while $C_{n \ge 1}$ are the azimuthal coefficients integrated over the final-state phase space in the same way as $C_0$ (see Eq.\eref{DY_distribution}).
The $R_{n0}$ moment has an immediate physical interpretation, being the average value of the cosine $\langle \cos \varphi \rangle$, and the $R_{n0}$ ones correspond to the higher moments $\langle \cos (n \varphi) \rangle$. 
The study of ratios between azimuthal correlations, $R_{nm} \equiv C_n/C_m = \langle \cos (n \varphi) \rangle / \langle \cos (m \varphi) \rangle$, was proposed in Refs.\tcite{Vera:2006un,Vera:2007kn}. Comparison of NLA predictions for azimuthal ratios in the Mueller--Navelet light-di-jet channel have shown a nice agreement with LHC data at $\sqrt{s} = 7$ TeV and for symmetric $p_T$-ranges of the two emitted jets\tcite{Ducloue:2013hia,Ducloue:2013bva,Caporale:2014gpa}.

\begin{figure*}[!t]
\centering
\includegraphics[scale=0.53,clip]{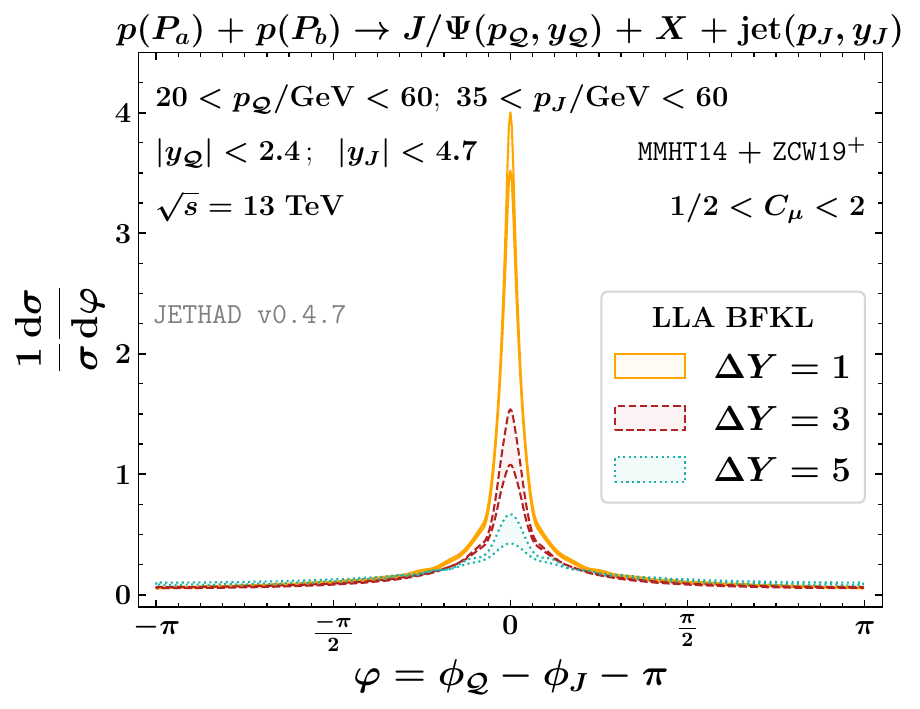}
\includegraphics[scale=0.53,clip]{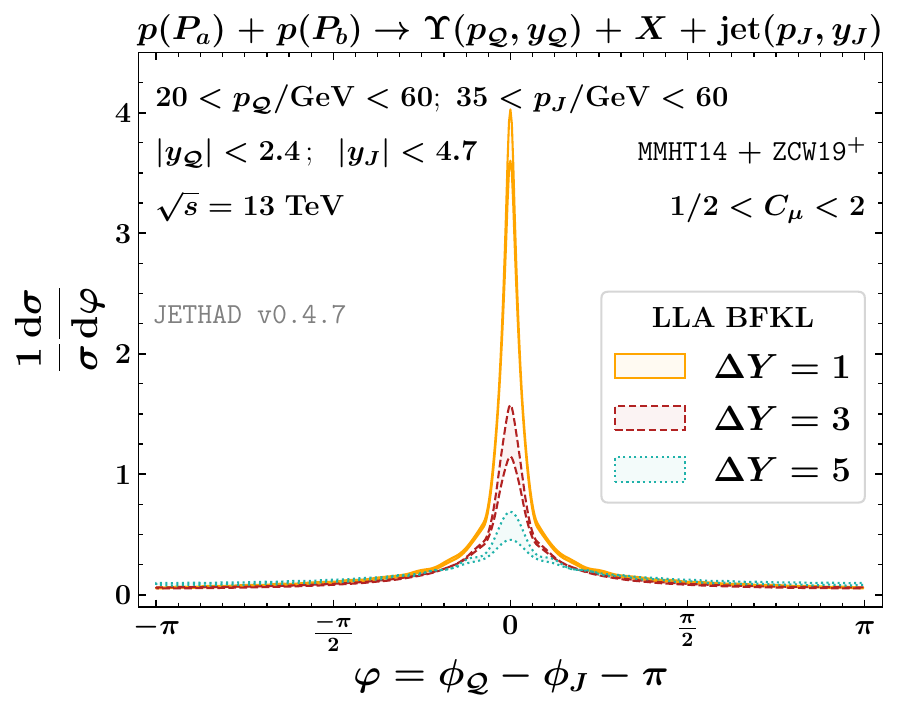}

\includegraphics[scale=0.53,clip]{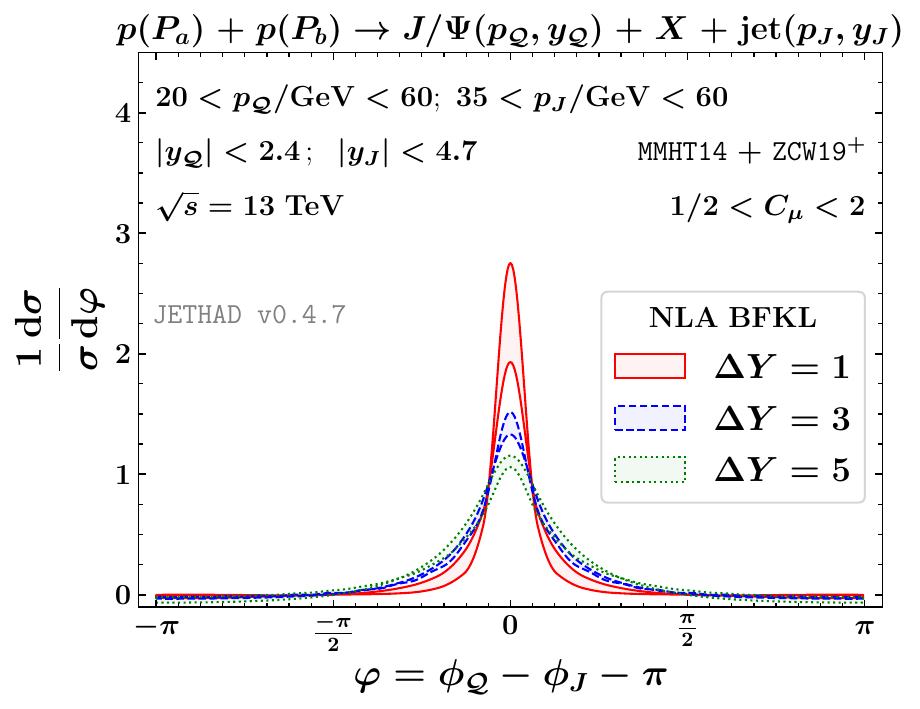}
\includegraphics[scale=0.53,clip]{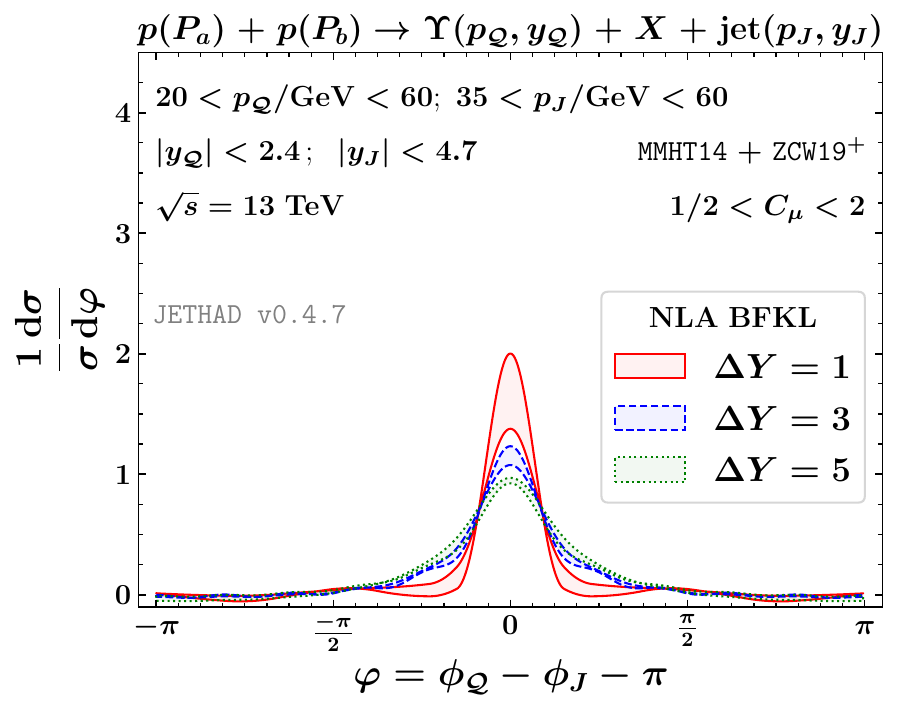}

\caption{LLA (upper panels) and NLA (lower panels) predictions for the $\varphi$-distribution in the $\JPsi$~$+$~jet (left) and in the $\Yps$~$+$~jet (right) channel, at $\sqrt{s} = 13$ TeV, and for three distinct values of $\DY$.
Quarkonium fragmentation is described in terms of {\tt ZCW19$^+$} functions. Text boxes inside panels exhibit final-state kinematic cuts. Uncertainty bands embody the combined effect of scale variation and phase-space multi-dimensional integration.}
\label{fig:phi_distribution}
\end{figure*}

As already mentioned, a major issue emerging in the NLA BFKL description of the Mueller--Navelet di-jet azimuthal ratios is the rise of instabilities of the high-energy series, which turned out to so be strong to prevent any possibility to perform realistic analyses at natural scales\tcite{Ducloue:2013hia,Caporale:2014gpa,Caporale:2015uva}. This phenomenon was later observed also in the inclusive semi-hard light
di-hadron\tcite{Celiberto:2017ptm,Celiberto:2020wpk} and hadron-jet production\tcite{Celiberto:2020wpk}. 
Recent studies on inclusive forward heavy-flavored hadrons have shown how the stabilization effect coming from the use of heavy-flavor VFNS FFs is quite strong where di-hadron final-state systems are considered, but it is less effective on hadron-plus-jet concurrent emissions\tcite{Celiberto:2021dzy,Celiberto:2021fdp}.

In this Section we focus on an alternative azimuthal-angle dependent observable, namely the $\varphi$-distribution, or simply azimuthal distribution
\begin{equation}
\label{dsigma_dphi}
 \frac{1}{\sigma}
 \frac{\drv \sigma}{\drv \varphi} = \frac{1}{2 \pi} \left\{ 1 + 2 \sum_{n = 1}^{\infty} \cos(n \varphi) \langle \cos(n \varphi) \rangle \right\}
 = \frac{1}{2 \pi} \left\{ 1 + 2 \sum_{n =1 }^{\infty} \cos(n \varphi) R_{n0} \right\} \; .
\end{equation}
Proposed for the first time in the context for Mueller--Navelet studies\tcite{Marquet:2007xx,Ducloue:2013hia}, the $\varphi$-distribution brings with it two main advantages. On the theoretical side, it collects the high-energy signals coming from all the $C_n$ coefficients, thus representing one of the most solid observables where to hunt for resummation effects. On the experimental side, since due to detector limitations the measured distributions cannot cover the whole $2 \pi$ azimuthal-angle range, a $\varphi$-dependent observable turns out to be much easier to be compared with data, rather than a $R_{nm}$ ratio. At the same time, due to the large number of $C_n$ coefficients required, numerical computations of Eq.\eref{dsigma_dphi} can be time consuming. We checked the numerical stability of our calculation by progressively raising the effective upper limit of the $n$-sum in Eq.\eref{dsigma_dphi}. An excellent numerical convergence was found at $n_{\rm max} = 20$.

In Fig.\tref{fig:phi_distribution} we present predictions for the azimuthal distribution as a function of $\varphi$ and for three distinct values of the rapidity interval, $\DY = 1, 3, 5$. Results for the $\JPsi$~$+$~jet hadroproduction are given in left panels, while the ones for the $\Yps$~$+$~jet hadroproduction are shown in right panels.
Results were obtained by making use of the {\tt ZCW19$^+$} set.
We adopted the same final-state kinematic cuts introduced in Section\tref{ssec:C0}.
A general feature of all the presented distributions is the presence of a clear peak at $\varphi = 0$. At this value the quarkonium and the jet are emitted back-to-back.
In all cases the height of the peak visibly diminishes when $\DY$ grows, while the distribution width slightly widens. This is distinctive signal of the onset of the high-energy dynamics. Indeed, at large values of $\DY$ the weight of gluons strongly ordered in rapidity predicted by BFKL increases, thus bringing to a reduction of  the azimuthal correlation between the quarkonium and the jet, so that the number of back-to-back events lowers.

Focusing on patterns for the $\varphi$-distribution in the $\JPsi$~$+$~jet channel, we observe that at $\DY = 1$ the LLA peak is much more pronounced than the corresponding NLA one, at $\DY = 3$ the LLA and NLA peaks are similar in height, and at $\DY = 5$ the LLA peak is beyond the NLA one.
This behavior as a straightforward explanation. The small-$\DY$ range stays at the limit of applicability of the BFKL resummation, since the low values of the partonic center-of-mass energies reduces the phase space for secondary-gluon emissions. This leads to a stronger discrepancy between LLA and NLA results. Conversely, in the moderate-$\DY$ regime the high-energy series shows a fair stability when NLA corrections are switched on. Finally, the large-$\DY$ territory is very sensitive to the BFKL dynamics, this reflecting in a stronger weight of the re-correlation effects, predicted by the NLA resummation, over pure LLA results.

Considering $\Yps$~$+$~jet production, we observe that LLA predictions are very close to the $\JPsi$~$+$~jet corresponding ones, whereas NLA results exhibit a different pattern. In particular, peaks are globally smaller, width are larger and uncertainty bands around peaks are very close to each other as $\DY$ increases.
This is a further indication that $\Yps$ emissions have a less stabilizing power on the high-energy resummation than $\JPsi$ ones.
At the same time, we remark that azimuthal distributions for both the considered channels feature a cleaner separation among results for different values of $\DY$ with respect to what happens for other reactions. As an example, error bands for the $\varphi$-distribution for the inclusive heavy-light di-jet production are definitely more overlapped, both at LLA and NLA~(see Fig.~9 of Ref.\tcite{Bolognino:2021mrc}).

The general outcome of results presented in this Section is the possibility of studying the azimuthal distribution for quarkonium-plus-jet processes around natural values of $\mu_R$ and $\mu_F$ scales. This observable can be easily measured at the LHC, thus offering the possibility of doing stringent tests of the high-energy resummation.

\section{Summary and Outlook}
\label{sec:conclusions}

We proposed the \emph{direct}  inclusive hadroproduction of a forward $^3S_1^{(1)}$ quarkonium, $\JPsi$ or $\Yps$, in association with a backward light jet in hybrid high-energy and collinear factorization.
We described the vector-meson hadronization in terms of a novel model of heavy-quark to quarkonium VFNS collinear FFs\tcite{Zheng:2019dfk}, suited for the study of quarkonium production at large transverse momentum, where the single-parton fragmentation mechanism is expected to prevail on short-distance $(Q \bar Q)$-pair production. We built two novel collinear FF sets, respectively named {\tt ZCW19} and {\tt ZCW19$^+$} functions. The first one was obtained trough the DGLAP evolution of a NRQCD input for the fragmentation of a heavy quark (charm or bottom) into a $^3S_1^{(1)}$ state at NLO\tcite{Zheng:2019dfk}. The second one combines the heavy-quark input\tcite{Zheng:2019dfk} with the gluon one, originally calculated in Ref.\tcite{Braaten:1993rw}. We believe that these functions can serve as useful tools for further phenomenological applications outside the high-energy domain, as well as for formal studies on the connection between the collinear factorization and the NRQCD effective formalism.

Motivated by recent studies on forward emissions of heavy-flavored hadrons in the same formalism\tcite{Celiberto:2021dzy,Celiberto:2021fdp}, we hunted for stabilizing effects of the high-energy resummation under higher-order corrections and energy-scale variation.
These effects are present and are even stronger when a forward $\JPsi$ is tagged rather than another charmed hadron, as the $\Lambda_c$ baryon. Such clues are less evident when $\Yps$ emissions are compared with bottomed-hadron ones. This is due to larger values of the $b$-quark mass that push the threshold for DGLAP evolution of the FFs closer to the energies provided by the considered kinematic regime, thus slightly worsening the validity of a pure VFNS description.
Future investigations are needed to gauge the dependence of our predictions on intrinsic features of the jet-emission description, as the jet selection function.

This work represents a further step in our ongoing program on heavy-flavored emissions, started from the analytic calculation of heavy-quark pair impact factors\tcite{Celiberto:2017nyx,Bolognino:2019ouc,Bolognino:2019yls}, and a first contact with the phenomenology of quarkonium production at high energies.
We plan to extend our analyses by considering quarkonium detections in wider kinematic ranges, as the ones accessible at new-generation colliding facilities, as the Electron-Ion Collider~(EIC)\tcite{Accardi:2012qut,AbdulKhalek:2021gbh,Khalek:2022bzd}, NICA-SPD\tcite{Arbuzov:2020cqg,Abazov:2021hku}, the High-Luminosity LHC\tcite{Chapon:2020heu,Amoroso:2022eow,Hentschinski:2022xnd},  the Forward Physics Facility~(FPF)\tcite{Anchordoqui:2021ghd,Feng:2022inv,Celiberto:2022rfj}, and the International Linear Collider~(ILC)\tcite{AlexanderAryshev:2022pkx}.
Here, our hybrid high-energy and collinear factorization could serve as an additional theoretical tool to perform quarkonium studies that cover a broad $p_T$-spectrum, with the aim of shedding light on the transition region from the short-distance $(Q \bar Q)$-pair production to the single-parton fragmentation mechanism.

\clearpage

\section*{Acknowledgments}

The authors are grateful to Alessandro~Papa and Dmitry~Yu.~Ivanov for a critical reading of the article and for useful suggestions.
The authors would like to express their gratitude to Jean-Philippe Lansberg for insightful conversations on the production mechanisms and phenomenology of quarkonia, and for encouragement.
We thank Valerio~Bertone and Hua-Sheng~Shao for inspiring discussions on the interplay between collinear factorization and NRQCD fragmentation.
The exciting conversations and the warm atmosphere of the \emph{Quarkonia As Tools} series of workshops inspired us in realizing the studies proposed in this work.
F.G.C. acknowledges support from the INFN/NINPHA project and thanks the Universit\`a degli Studi di Pavia for the warm hospitality.
M.F. acknowledges support from the INFN/QFT@COL\-LI\-DERS project.
%

\bibliographystyle{apsrev}
\bibliography{references}

\end{document}